\documentclass[11pt,reqno]{amsart}

\usepackage{amsmath}
\usepackage{amssymb}
\usepackage{amsfonts}
\usepackage{amsthm}
\usepackage{a4wide}
\usepackage{bbm}
\usepackage{graphics}
\usepackage{color}
\usepackage[sans]{dsfont}

\newtheorem{theorem}{Theorem}[section]
\newtheorem{lemma}[theorem]{Lemma}
\newtheorem{proposition}[theorem]{Proposition}
\newtheorem{corollary}[theorem]{Corollary}

\newtheorem{remark}[theorem]{Remark}

\numberwithin{equation}{section}

\newcommand{\ii}{\mathrm{i}}
\renewcommand{\d}{\mathrm{d}}

\newcommand{\D}{\mathcal{D}}

\def\C{{\mathbb C}}

\def\N{{\mathbb N}}
\def\R{{\mathbb R}}

\def\CO{{\mathcal O}}
\def\CR{{\mathcal R}}

\newcommand{\cH}{\mathcal{H}}
\newcommand{\cF}{\mathcal{F}}

\newcommand{\ve}{{\varepsilon}}

\newcommand{\lam}{{\lambda}}           

\newcommand{\Om}{\Omega}                

\newcommand{\fh}{\mathfrak{h}}

\def\<{\langle}
\def\>{\rangle}

\newcommand{\ra}{\rightarrow}

\newcommand{\one}{\mathds{1}}

\newcommand{\im}{\operatorname{Im}}
\newcommand{\re}{\operatorname{Re}}
\newcommand{\ddim}{\operatorname{d\:\!Im}}
\newcommand{\ddre}{\operatorname{d\:\!Re}}
\newcommand{\supp}{\operatorname{supp}}
\newcommand{\dive}{\operatorname{div}}

\newcommand{\slim}{\mathop{\text{\rm{s-lim}}}}

\newcommand{\DETAILS}[1]{}

\newcommand{\fract}[2]{\genfrac{}{}{0pt}{}{\scriptstyle #1}{\scriptstyle #2}}

\title{Maximal velocity  of photons in non-relativistic QED}

\author[J.-F. Bony]{Jean-Fran\c{c}ois Bony}
\address[J.-F. Bony]{Institut de Math{\'e}matiques de Bordeaux \\
UMR-CNRS 5251, Universit{\'e} de Bordeaux 1 \\
351 cours de la lib{\'e}ration, 33405 Talence Cedex, France}
\email{bony@math.u-bordeaux1.fr}
\author[J. Faupin]{J{\'e}r{\'e}my Faupin}
\address[J. Faupin]{Institut de Math{\'e}matiques de Bordeaux \\
UMR-CNRS 5251, Universit{\'e} de Bordeaux 1 \\
351 cours de la lib{\'e}ration, 33405 Talence Cedex, France}
\email{jeremy.faupin@math.u-bordeaux1.fr}
\author[I. M. Sigal]{Israel Michael Sigal}
\address[I. M. Sigal]{Department of Mathematics \\
University of Toronto \\
40 St. George Street, Bahen Centre, Toronto, ON M5S 2E4, Canada}
\email{im.sigal@utoronto.ca}

\begin{document}

\begin{abstract}
We consider the problem of propagation of photons in the quantum theory of non-relativistic matter coupled to electromagnetic radiation, which is, presently, the only consistent quantum theory of matter and radiation. Assuming that the matter system is in a localized state (i.e for energies below the ionization threshold), we show that the probability to find photons at time $t$ at the distance greater than $\mathrm{c} t$, where $\mathrm{c}$ is the speed of light, vanishes as $t\ra \infty$ as an inverse power of $t$.
\end{abstract}

\maketitle

\section{Introduction}

One of the key postulates in the theory of relativity is that the speed of light is constant and the same in all inertial reference frames. This postulate, verified to begin with experimentally, can also be easily checked theoretically for propagation of disturbances in the free Maxwell equations. However, one would like to show it for the physical model of matter interacting with electromagnetic radiation. To have a sensible model, one would have to consider both matter and radiation as quantum. This, in turn, requires reformulation of the problem in terms of quantum probabilities. The latter are given through localization observables for photons. We define it below. Now we  proceed to  the model of quantum matter interacting with (quantum) radiation. (By radiation we always mean the electromagnetic radiation.) In what follows we use the units in which the speed of light and the Planck constant divided by $2 \pi$ are $1$.

Presently, the only mathematically well-defined such a model, which is in a good agreement with experiments, is the one in which matter is treated non-relativistically. In this model, the state space of the total system is given by $\cH=\cH_{p}\otimes\cH_{f}$, where $\cH_{p}$ is the state space of the particles, say $\cH_{p} = \mathrm{L}^2(\R^{3n})$, and $\cH_{f}$ is the state spaces of photons (i.e. of the quantized electromagnetic field), defined as the bosonic (symmetric) Fock space, $\cF$, over the one-photon space $\fh$ (see Appendix \ref{s8} for the definition of $\mathcal{F}$). In the momentum representation, $\fh$ is the $\mathrm{L}^2$-space, $\mathrm{L}^2_{\textrm{transv}}( \mathbb{R}^3 ; \C^3 )$, of complex vector fields $f:\R^3 \ra \C^3$, satisfying $k\cdot f(k)=0$. By choosing orthonormal vector fields $\ve_\lam(k):\R^3 \ra \C^3,\ \lam=1, 2$, satisfying $k\cdot \ve_\lam(k)=0$ ($\ve_\lam(k),\ \lam=1, 2$, are called the polarization vectors), we identify $\fh$ with the space $\mathrm{L}^2( \mathbb{R}^3 ; \C^2)$ of square integrable functions of photon momentum $k\in\R^3$ and polarization index $\lam=1, 2$.

The dynamics of the system is described by the Schr\"odinger equation,
\begin{equation}
\ii \partial_t \psi_{t} = H \psi_{t},
\end{equation}
on the state space $\cH=\cH_{p}\otimes\cH_{f}$, with the standard quantum Hamiltonian (see \cite{Fe32_01,PaFi38_01})
\begin{equation*}
H = \sum \limits_{j=1}^n \frac{1}{2 m_j} \big( - \ii \nabla_{x_j} - \varepsilon_{j} A_\kappa ( x_j ) \big)^2 + V (x) + H_f .
\end{equation*}
Here, $m_j$ and $x_j$, $j=1, \ldots , n$, are the (`bare') particle masses and the particle positions, $V(x)$, $x = ( x_1 , \ldots , x_n )$, is the total potential affecting the particles and $\varepsilon_{j}$ are coupling constants related to the particle charges. Moreover, $A_\kappa : = \check \kappa *A$, where $A(y)$ is the \textit{quantized vector potential} in the Coulomb gauge ($\dive A(y)=0$), describing the quantized electromagnetic field, and given by
\begin{equation} \label{a73}
A_\kappa(y)=\sum_{\lambda=1,2} \int \varepsilon_\lambda (k) \big( e^{\ii k y} a_\lambda (k) + e^{- \ii k y} a_\lambda^* (k) \big) \kappa (k) \frac{\d k}{\sqrt{2 \vert k \vert}} ,
\end{equation}
where $\kappa \in \mathrm{C}^{\infty}_{0} ( \R^{3} )$ is an \textit{ultraviolet cut-off}. The operator $H_{f}$ is the quantum Hamiltonian of the quantized electromagnetic field, describing the dynamics of the latter,
\begin{equation} \label{Hf}
H_{f} = \sum_{\lambda = 1 , 2} \int \omega ( k ) a_{\lambda}^* (k) a_{\lambda} (k) \, \d k ,
\end{equation}
where $\omega ( k ) = \vert k \vert$ is the dispersion relation. The integrals without indication of the domain of integration are taken over entire $\mathbb{R}^3$. Above, $\lambda$ is the polarization, $a_\lambda(k)$ and $a_\lambda^*(k)$ are annihilation and creation operators
acting on the Fock space $\cH_{f} = \cF$ (see Appendix \ref{s8} for the definition of annihilation and creation operators).

Assuming for simplicity that our matter consists of electrons and nuclei and that the nuclei are infinitely heavy and therefore are manifested through the interactions only (put differently, the molecules are treated in the Born--Oppenheimer approximation), one arrives at the operator $H$ with the coupling constants $\varepsilon_{j} : = \alpha^{1 / 2}$, where $\alpha =\frac{e^2}{4\pi \hbar \mathrm{c}}\approx \frac{1}{137}$ is the fine-structure constant. After that one can relax the conditions on the potentials $V(x)$ allowing say general many-body ones (see \cite{GuSi11_01} for a discussions of the Hamiltonian $H$).
Since the structure of the particles system is immaterial for us, to keep notation as simple as possible, we consider a single particle in an external potential, $V(x)$, coupled to the quantized electromagnetic field. Furthermore, since our results do not depend on the value of $\alpha$, we absorb it into the ultraviolet cut-off $\kappa$. In this case, the state space of such a system is $\cH=\mathrm{L}^2( \mathbb{R}^3) \otimes\cF=\mathrm{L}^2( \mathbb{R}^3 ; \mathcal{F} )$ and the standard Hamiltonian operator acting on $\mathrm{L}^2( \mathbb{R}^3 ; \mathcal{F} )$ is given by (we omit the subindex $\kappa$ in $A(x)$)
\begin{equation}
H : = \big( p + A(x) \big)^2 + H_f + V (x),
\end{equation}
with the notation $p = - \ii \nabla_{x}$, the particle momentum operator. We assume that $V$ is real valued and infinitesimally bounded with respect to $p ^{2}$.

Our goal is to show that photons departing a bound particle system, say an atom or a molecule, move away from it with a speed not higher than the speed of light. Let $\d \Gamma (A)$ denote the lifting of a one-photon operator $A$ to the photon Fock space (and then to the Hilbert space of the total system), $y := \ii \nabla_k$ be the operator on $\mathrm{L}^2( \mathbb{R}^3 ; \C^2)$, canonically conjugate to the photon momentum $k$ and let $\one_\Om ( y )$ denote the characteristic function of a subset $\Om$ of $\R^3$. To test the photon localization, we define the observables $\d \Gamma ( \one_\Om ( y ) )$, which can be interpreted as giving the number of photons in Borel sets $\Om\subset \R^3$. These observables are closely related to those used in \cite{FrGrSc02_01,Ge02_01,LiLo03_01} and are consistent
with a theoretical description of detection of photons (usually via the photoelectric effect, see e.g. \cite{MaWo95_01}).\footnote{The issue of localizability of photons is a tricky one and has been intensely discussed in the literature since the 1930 and 1932 papers by Landau and Peierls \cite{LaPe30_01} and Pauli \cite{Pa64_01} (see also a review in \cite{Ke05_01}). A set of axioms for localization observables was proposed by Newton and Wigner \cite{NeWi49_01} and Wightman \cite{Wi62_01} and further generalized by Jauch and Peron \cite{JaPi67_01}. Localization observables for massless particles satisfying the Jauch--Peron version of the Wightman axioms were constructed by Amrein \cite{Am69_01}.} The fact that they depend on the choice of polarization vector fields, $\ve_\lam(k),\ \lam=1, 2,$ is not an impediment here as our results imply analogous results for e.g. similarly constructed observables\footnote{These observables are similar to those introduced by Mandel \cite{Ma66_01}. Since polarization vector fields are not smooth, using them to reduce the results from one set of localization observables to another would limit the possible time decay. However, these vector fields can be avoided by using the approach of \cite{LiLo04_01}.} based on the space $\mathrm{L}^2_{\textrm{transv}}( \mathbb{R}^3 ; \C^3 )$ instead of $\mathrm{L}^2( \mathbb{R}^3 ; \C^2)$, or localization observables constructed by Amrein \cite{Am69_01}. (Both latter observables are also covariant, $T_g\d \Gamma ( \one_\Om ( y ) ) T_g^{-1} = \d \Gamma ( \one_{g^{-1}\Om} ( y ) )$, under rigid motions, $g$, of $\R^3$, generated by one particle transformations, $f(y) \ra f(g^{-1}y)$, as one would like to have for localization observables.)

With the definition of localization observables given, we say that photons propagate with speed $\leq \mathrm{c}'$ if for any initial condition $\psi_0$ and for any $\mathrm{c} > \mathrm{c}'$, the state, $\psi_t$, of the system at time $t$, satisfies the estimate
\begin{equation*}
\big\Vert \d \Gamma \big( F ( \vert y \vert \geq \mathrm{c} t ) \big)^{\frac{1}{2}} \psi_t \big\Vert \longrightarrow 0 \qquad \text{as } t \rightarrow \infty ,
\end{equation*}
for any function $F ( s \geq 1 )$ supported in the domain $\{ s \geq 1 \}$. Similarly, one can define the propagation with speed $\ge \mathrm{c}'$. As with any other quantum models, this definition allows for a non-zero probability that photons propagate with arbitrary high speed. However, as estimates of such probabilities for massive free relativistic particles show (see \cite{Ru81_01}), these events (as with the problem of reversibility) have so low probabilities as to make them undetectable.

To formulate our result, we let $\Sigma$ denote the ionization threshold defined by
\begin{equation*}
\Sigma : = \lim_{R \to \infty} \inf_{\fract{\varphi \in D_{R}}{\Vert \varphi \Vert = 1}} \< \varphi , H \varphi \> ,
\end{equation*}
where $D_{R} = \{ \varphi \in \D ( H ) ; \ \varphi ( x ) = 0 \text{ if }\vert x \vert < R \}$ (see \cite{Gr04_01}). Let $f \in \mathrm{C}_0^\infty( \mathbb{R} ; [0,1] )$ be such that $\supp (f) \subset [1,2]$ and define $F(s) = \int_{- \infty}^s f ( \tau ) \, \d \tau$. We will localize the photon position using the following operator
\begin{equation} \label{F}
F (\vert y \vert \ge \mathrm{c}t ) = F ( \vert v \vert \ge 1 ) := F ( \vert v \vert),
\end{equation}
where $v := y/\mathrm{c}t$. Our main result is the following

\begin{theorem}\sl \label{t1}
Let $F$ be as above, $\chi \in \mathrm{C}_0^\infty( (-\infty,\Sigma))$ and $\mathrm{c} > 1$. For all $u \in \D( \d \Gamma( \< y \> )^{\frac{1}{2}} )$, we have
\begin{equation*}
\Big\Vert \d \Gamma \big( F ( \vert y \vert  \geq \mathrm{c} t  ) \big)^{\frac{1}{2}} e^{-\ii t H} \chi( H ) u \Big\Vert \lesssim t^{- \gamma} \big\Vert \big( \d \Gamma ( \< y \> ) + 1 \big)^{\frac{1}{2}} u \big\Vert ,
\end{equation*}
where
\begin{equation} \label{a59}
\gamma < \min \Big( \frac{1}{2} \Big( 1-  \frac{1}{ \mathrm{c} } \Big) , \frac{1}{10} \Big) .
\end{equation}
\end{theorem}

Thus $e^{-\ii t H} \chi( H ) u$ is supported asymptotically in the set $\vert y \vert  \le \mathrm{c} t$. In other words, photons do not propagate faster than the speed of light.

The estimate of Theorem 1.1 is usually called a \emph{strong
propagation estimate} in the literature (see \cite{DeGe97_01,SiSo88_01}). In order to prove it, we first
need to `improve' the infrared behavior of the electron-photons
interaction given by \eqref{a73}, which can be done, as usual, by
performing a \emph{Pauli--Fierz transformation}. For technical
convenience, we use a generalized Pauli--Fierz transformation as in
\cite{Si09_01}. Next, we employ the method of propagation observables by
constructing a positive, unbounded observable, whose Heisenberg
derivative is negative (up to integrable remainder terms). In our
proof, the required estimates on the remainder terms are obtained
thanks to Hardy's inequality in $\mathbb{R}^3$, together with a
suitable control of the growth of $\d \Gamma( |k|^{-\delta} )$ along
the evolution, for some $0 \le \delta \le 1$.

For massive Pauli--Fierz Hamiltonians (that is with a dispersion
relation of the form $\omega(k) = \sqrt{ k^2 + m^2}$, $m>0$), a weak
version of the maximal velocity estimate is derived in
\cite{DeGe99_01} (see also \cite{FrGrSc02_01} for a different weak maximal velocity
estimate). Compared to \cite{DeGe99_01}, the main difficulty we encounter is
that, in our case, the number of photons operator is \emph{not}
relatively bounded with respect to the Hamiltonian. It is presently
not known whether or not the number of photons remains bounded along
the evolution (see, however, the recent paper \cite{DeKu11_01} for the case
of massless spin-boson model). Another difficulty here is due to the lack of smoothness of the relativistic dispersion relation $\omega(k) = \vert k \vert$ at the origin.

Our paper is organized as follows. In Section \ref{s2}, we introduce a
generalized Pauli--Fierz transformation and prove our main theorem.
Various ingredients of the proof of Theorem \ref{t1} are deferred to the next sections. In
Section \ref{s3}, we estimate interaction terms. Section \ref{s4} is devoted to the
estimate of the growth of $\d \Gamma( |k|^{-\delta} )$ along the
evolution. In Section \ref{s5}, we control remainder terms by estimating
some commutators. A few standard estimates are gathered in Appendix \ref{s6},
domain questions are discussed in Appendix \ref{s7}, and finally, for the
convenience of the reader, standard definitions of operators in Fock
space are recalled in Appendix \ref{s8}.

\section{Proof of Theorem \ref{t1}} \label{s2}

To prove Theorem \ref{t1}, we use the generalized Pauli--Fierz transformation (see \cite{Si09_01}) defined as follows. For any $h \in \mathrm{L}^2( \mathbb{R}^3 ; \C^2)$, we define the operator-valued  field
\begin{equation}
\Phi( h ) := \frac{1}{ \sqrt{2} } ( a^*(h) + a(h) ) .
\end{equation}
Using it, we can write
\begin{equation}
A( x )  = \Phi ( g_{x} ), \qquad g_{x} ( k , \lambda ) := \frac{ \kappa(k) }{ |k|^{\frac{1}{2}} } \varepsilon_\lambda(k) e^{ \ii k \cdot x }.
\end{equation}
Let $\varphi \in \mathrm{C}^\infty ( \mathbb{R} ; \mathbb{R} )$ be a non-decreasing function such that $\varphi(r) = r$ if $|r| \leq 1/2$ and $| \varphi(r) | = 1$ if $|r| \geq 1$. For $0 < \mu < 1 / 2$, we define the function
\begin{equation*}
q_{x} (k,\lambda) := \frac{ \kappa(k) }{ |k|^{\frac{1}{2} + \mu } } \varphi( |k|^\mu \varepsilon_\lambda(k) \cdot x ),
\end{equation*}
and the unitary operator
\begin{equation*}
\mathcal{U} := e^{- \ii \Phi ( q_{x} )} ,
\end{equation*}
on $\mathrm{L}^2 ( \R^3 ; \cF )$. We also introduce the Pauli--Fierz transformed Hamiltonian $\widetilde{H}$ by $\widetilde{H} := \mathcal{U} H \mathcal{U}^*$. We compute
\begin{equation*}
\widetilde{H}  = \big( p + \widetilde{A} (x) \big)^2 + E(x) + H_f + \widetilde{V} (x) ,
\end{equation*}
where
\begin{gather*}
\widetilde{A} (x) := \Phi( \widetilde{g}_{x} ), \qquad \widetilde{g}_{x} ( k,\lambda) := g_{x} ( k,\lambda) - \nabla_x q_{x} ( k,\lambda), \\
E(x) := \Phi( e_{x} ), \qquad e_{x} ( k,\lambda) := \ii |k| q_{x} ( k,\lambda), \\
\widetilde{V} (x) :=V(x) + \frac{1}{2} \sum_{\lambda=1,2} \int_{ \mathbb{R}^3 } | k | | q_{x} ( k , \lambda ) |^2 \d k.
\end{gather*}
The generalized Pauli--Fierz transformation is technically convenient since the operator $\widetilde{H}$ is self-adjoint with domain $\D( \widetilde{H} ) = \D( H ) = \D ( p^2 + H_f )$ (see Theorem \ref{a25} in Appendix \ref{s7}).

The coupling functions $q_{x} ( k , \lambda )$, $\widetilde{g}_{x} ( k , \lambda )$ and $e_{x} ( k , \lambda )$ satisfy the estimates
\begin{gather}
| \partial_{k}^m q_x(k,\lambda) \big | \lesssim \kappa_m(k)  |k|^{- \frac{1}{2} - | m |}\< x \>^{1 + |m|},   \label{a56} \\
|\partial_{k}^m  \widetilde{g}_x(k,\lambda) | \lesssim  \kappa_m(k) \vert k \vert^{\frac{1}{2} - \vert m \vert} \< x \>^{\frac{1}{\mu} + \vert m \vert} ,     \label{a57}  \\
|\partial_{k}^m  e_x(k,\lambda) | \lesssim \kappa_m(k) |k|^{\frac{1}{2} - \vert m \vert} \< x \>^{1 + \vert m \vert},   \label{a58}
\end{gather}
where $\kappa_m(k) \ge 0$ is compactly supported and bounds $ \kappa(k) $ and all its derivatives up to the order $|m|$. These estimates will play an important role in our analysis. \eqref{a56} and \eqref{a58} follow directly from the definition of $q_{x}$ and $e_{x}$. To obtain \eqref{a57} for $m=0$, we use
\begin{align*}
| \widetilde{g}_x(k,\lambda) | &= \frac{ | \kappa(k) | }{ |k|^{\frac{1}{2}} } \big | e^{ \ii k \cdot x } - \varphi^{\prime} ( |k|^\mu \varepsilon_\lambda(k) \cdot x ) \big |    \\
&\leq \frac{| \kappa(k) |}{|k|^{\frac{1}{2}}}  \Big( \big| e^{\ii k \cdot x} -1 \big| + \big| 1 - \varphi^{\prime} ( |k|^\mu \varepsilon_\lambda(k) \cdot x ) \big| \Big) ,
\end{align*}
and the estimates $|e^{ \ii k \cdot x }-1|\lesssim |k| |x|$ and $| 1 - \varphi^{\prime} ( |k|^\mu \varepsilon_\lambda(k) \cdot x ) |\lesssim ( |k|^\mu |x|)^r$ for all $r > 0$. The latter is implied by the property that $1 - \varphi^{\prime} ( |k|^\mu \varepsilon_\lambda(k) \cdot x )=0$ for $|k|^\mu \varepsilon_\lambda(k) \cdot x\le \frac{1}{2}$. Choosing $r=1/\mu$, we arrive at \eqref{a57} for $m=0$. The case of $|m|>0$ is treated similarly.

We shall prove

\begin{theorem}\sl \label{t2}
Let $F$ be as in \eqref{F}, $\chi \in \mathrm{C}_0^\infty ( ( - \infty , \Sigma ) )$ and $\mathrm{c} > 1$. For all parameters $\beta , \gamma , \delta$ such that
\begin{gather}
0 \leq \beta < \delta < 1 ,     \label{a49}  \\
0 \leq \gamma < \min \Big( \Big( 1 - \frac{1}{\mathrm{c}} \Big) \beta , \frac{3 \delta - 2}{10} \Big) ,   \label{a50}
\end{gather}
we have, for $u \in \D \big( \d \Gamma ( \vert k \vert^{- \delta} )^{\frac{1}{2}} \big) \cap \D \big( \d \Gamma ( \vert y \vert^{2 \beta} )^{\frac{1}{2}} \big)$,
\begin{equation}  \label{a55}
\Big\Vert \d \Gamma \big( F ( \vert y \vert \geq \mathrm{c} t ) \big)^{\frac{1}{2}} e^{- \ii t \widetilde{H}} \chi( \widetilde{H} ) u \Big\Vert \lesssim t^{- \gamma} \Big\Vert \big( \d \Gamma \big( \vert k \vert ^{- \delta} + \vert y \vert^{2 \beta} \big) + 1 \big)^{\frac{1}{2}} u \Big\Vert .
\end{equation}
\end{theorem}

We first verify that Theorem \ref{t2} implies Theorem \ref{t1} and next proceed to the proof of Theorem \ref{t2}.

\begin{proof}[Proof of Theorem \ref{t1}]
For $\gamma$ as in \eqref{a59}, we fix $\beta$ and $\delta$ satisfying \eqref{a50} and $0 \leq 2 \beta < \delta < 1$. Let $\widehat{\chi} \in \mathrm{C}_0^\infty( (-\infty , \Sigma ) )$ be such that $\chi \widehat{\chi} = \chi$. We set $u_{t} : = e^{- \ii t H} \chi ( H ) u$ and $\widehat{u} := \widehat{\chi} ( H ) u$. Using the Pauli--Fierz transformation $\mathcal{U}$, we write
\begin{equation*}
\Big\Vert \d \Gamma ( F ( \vert v \vert ) )^{\frac{1}{2}} u_t \Big\Vert^2 = \Big\< e^{-\ii t \widetilde{H}} \chi( \widetilde{H} )\mathcal{U} \widehat{u} , \mathcal{U} \d \Gamma ( F ( \vert v \vert ) ) \mathcal{U}^* e^{-\ii t \widetilde{H}} \chi( \widetilde{H} )\mathcal{U} \widehat{u} \Big\>.
\end{equation*}
We compute
\begin{equation} \label{a60}
\mathcal{U} \d \Gamma ( F ( \vert v \vert ) ) \mathcal{U}^* = \d \Gamma ( F ( \vert v \vert ) ) + \Phi ( \ii F ( \vert v \vert ) q_{x} ) - \frac{1}{2} \re \big\< F ( \vert v \vert ) q_{x} , q_{x} \big\> .
\end{equation}
Using Corollary \ref{a11} and Theorem \ref{a23}, we can estimate the second term given by \eqref{a60} as
\begin{align}
\Big\vert \Big\< e^{- \ii t \widetilde{H}} \chi( \widetilde{H} ) \mathcal{U} & \widehat{u} , \Phi ( \ii F ( \vert v \vert ) q_{x} ) e^{-\ii t \widetilde{H}} \chi( \widetilde{H} )\mathcal{U} \widehat{u} \Big\> \Big\vert    \nonumber \\
&\lesssim \Big\Vert \Phi ( \ii F ( \vert v \vert ) q_{x} ) \< x \>^{-\tau_1} ( H_f + 1 )^{-\frac{1}{2}} \Big\Vert \Big\Vert (H_f+1)^{\frac{1}{2}} \< x \>^{\tau_1} \chi( \widetilde{H} ) \Big\Vert \Vert \widehat{u} \Vert^2    \nonumber \\
& \lesssim t^{-d_1} \Vert u \Vert^2,
\end{align}
with $0 \leq d_1 < 1/2$ and $\tau_1 = 3 / 2 + d_1$. Similarly, using Lemma \ref{a3} and Theorem \ref{a23}, the last term given by \eqref{a60} is estimated as
\begin{align}
\Big\vert \Big\< \chi( \widetilde{H} ) e^{-\ii t \widetilde{H}} \mathcal{U} & \widehat{u} , \re \big\< F ( \vert v \vert ) q_{x} , q_{x} \big\> \chi( \widetilde{H} ) e^{-\ii t \widetilde{H}} \mathcal{U} \widehat{u} \Big\> \Big\vert    \nonumber \\
&\lesssim \big\Vert F ( \vert v \vert ) q_{x} ( k , \lambda ) \< x \>^{-\tau_2} \big\Vert \big\Vert \< x \>^{\tau_2} \chi( \widetilde{H} ) \big\Vert \Vert\widehat{u}\Vert^2 \lesssim t^{-d_2} \Vert u \Vert^2,
\end{align}
with $0 \leq d_2 < 1$ and $\tau_2 = 1 + d_2$.

Now, by Theorem \ref{t2}, we have
\begin{align}
\Big\< \chi( \widetilde{H} ) e^{-\ii t \widetilde{H}} \mathcal{U} & \widehat{u} , \d \Gamma ( F ( \vert v \vert ) ) \chi( \widetilde{H} ) e^{-\ii t \widetilde{H}} \mathcal{U} \widehat{u} \Big\>       \nonumber \\
&\lesssim t^{-2\gamma} \Big\< \widehat{u} , \mathcal{U}^* \big( \d \Gamma ( \vert k \vert^{- \delta} ) + \d \Gamma ( \vert y \vert^{2\beta} ) + 1 \big) \mathcal{U} \widehat{u} \Big\> .
\end{align}
Therefore it remains to show that
\begin{align*}
\Big\< \widehat{u} , \mathcal{U}^* \big( \d \Gamma ( \vert k \vert^{- \delta} ) + \d \Gamma( \vert y \vert^{2\beta} ) + 1 \big) \mathcal{U} \widehat{u} \Big\> \lesssim \big\< u , \big( \d \Gamma ( \< y \> ) + 1 \big) u \big\> .
\end{align*}
We can compute as above
\begin{align}
\mathcal{U}^* \big( \d \Gamma ( \vert k \vert^{- \delta} ) + \d \Gamma( \vert y \vert^{2\beta} ) + 1 \big) \mathcal{U} ={}& \big( \d \Gamma ( \vert k \vert^{- \delta} ) + \d \Gamma( \vert y \vert^{2\beta} ) + 1 \big) - \Phi \big( \ii (\vert k \vert^{- \delta} + \vert y \vert^{2\beta} ) q_{x} \big)    \nonumber \\
&+ \frac{1}{2} \re \big\< ( \vert k \vert^{- \delta} + \vert y \vert^{2\beta} ) q_{x} , q_{x} \big\>. \label{a61}
\end{align}
Since $0 \leq 2 \beta \leq \delta$, Hardy's inequality (see Lemma \ref{a24}) together with Lemma \ref{a21} imply that
\begin{align*}
\Big\Vert \big( \d \Gamma ( \< y \>^{\delta} ) + 1 \big)^{-\frac{1}{2}} \big( \d \Gamma ( \vert k \vert^{- \delta} ) + \d \Gamma( \vert y \vert^{2\beta} ) + 1 \big) \big( \d \Gamma ( \< y \>^{\delta} ) + 1 \big)^{-\frac{1}{2}} \Big\Vert \lesssim 1 .
\end{align*}
Besides, using \eqref{a56}, $\delta < 1$ and Lemma \ref{a7}, one can estimate
\begin{gather*}
\Big\Vert \Phi \big( \ii \vert k \vert^{- \delta} q_{x} \big) ( N + 1 )^{-\frac{1}{2}} \< x \>^{- 1} \Big\Vert \lesssim \sup_{x \in \mathbb{R}^3} \big\Vert \vert k \vert^{- \delta} q_{x} ( k , \lambda ) \< x \>^{- 1} \big\Vert_{\mathrm{L}^2( \mathbb{R}^3 ; \C^{2} )} \lesssim 1 ,    \\
\big\Vert \big\< \vert k \vert^{- \delta} q_{x} , q_{x} \big\> \< x \>^{- 2} \big\Vert \lesssim \sup_{x \in \mathbb{R}^3} \big\Vert \vert k \vert^{- \frac{\delta}{2}} q_{x} ( k , \lambda ) \< x \>^{- 1} \big\Vert^{2}_{\mathrm{L}^2( \mathbb{R}^3 ; \C^{2} )} \lesssim 1 .
\end{gather*}
Similarly, by Lemma \ref{a3} (with $t=1$) and Lemma \ref{a7}, we have
\begin{gather*}
\Big\Vert \Phi \big( \ii \vert y \vert^{2 \beta} q_{x} \big) ( N + 1 )^{-\frac{1}{2}} \< x \>^{- 2} \Big\Vert \lesssim \sup_{x \in \mathbb{R}^3} \big\Vert \vert y \vert^{2 \beta} q_{x} ( k , \lambda ) \< x \>^{- 2} \big\Vert_{\mathrm{L}^2( \mathbb{R}^3 ; \C^{2} )} \lesssim 1 , \\
\big\Vert \big\< \vert y \vert^{2 \beta} q_{x} , q_{x} \big\> \< x \>^{- 3} \big\Vert \lesssim \sup_{x \in \mathbb{R}^3} \big\Vert \vert y \vert^{\beta} q_{x} ( k , \lambda ) \< x \>^{-\frac{3}{2}} \big\Vert^{2}_{\mathrm{L}^2( \mathbb{R}^3 ; \C^{2} )} \lesssim 1 ,
\end{gather*}
since $0 < \beta < 1/2$. Combining \eqref{a61}, the previous estimates and an interpolation argument, we obtain
\begin{equation*}
\Big\< \widehat{u} , \mathcal{U}^* \big( \d \Gamma ( \vert k \vert^{- \delta} ) + \d \Gamma( \vert y \vert^{2\beta} ) + 1 \big) \mathcal{U} \widehat{u} \Big\> \lesssim \big\< \widehat{u} , \big( \d \Gamma ( \< y \>^{\delta} ) + N + \< x \>^{6} + 1 \big) \widehat{u} \big\> .
\end{equation*}
To conclude, it suffices to use that
\begin{equation*}
\Big\Vert \d \Gamma ( \< y \>^{\delta} )^{\frac{1}{2}} \widehat{\chi} ( H ) \big( \d \Gamma ( \< y \> ) + 1 \big)^{- \frac{1}{2}} \Big\Vert \lesssim 1 ,
\end{equation*}
by Proposition \ref{a26}, together with
\begin{equation*}
\Big\Vert N^{\frac 12} \widehat{\chi} ( H ) \big( \d \Gamma ( \< y \> ) + 1 \big)^{-\frac{1}{2}} \Big\Vert \lesssim 1 ,
\end{equation*}
by Lemma \ref{a29}, and
\begin{align*}
\big\Vert \< x \>^{6} \widehat{u} \big\Vert \lesssim \Vert u \Vert ,
\end{align*}
by Theorem \ref{a23}.
\end{proof}

\begin{proof}[Proof of Theorem \ref{t2}]
We use the method of propagation observables by constructing a family of operators  $\Phi_t$ (called a \textit{propagation observable}) such that on one hand an appropriate bound on $\<  \widetilde{u}_t, \Phi_t  \widetilde{u}_t  \>$, where $\widetilde{u}_t = e^{-\ii t \widetilde{H}} \chi( \widetilde{H} ) u$, implies (a part of) the statement of the theorem and, on the other hand, $\Phi_t$ satisfies a differential inequality which implies this bound. Fix $\beta , \gamma , \delta$ satisfying \eqref{a49}--\eqref{a50}. We set
\begin{equation*}
J_\beta ( s ) : = s^\beta F ( s^{\frac{1}{2}} ) \in \mathrm{C}^{\infty} ( \R ).
\end{equation*}
The family  $\Phi_t$  is defined,  as a quadratic form on $\chi ( \widetilde{H} ) \D ( \d \Gamma ( \< y \>^{\beta} ) )$, by
\begin{align*}
\Phi_t := t^{2\gamma} \d \Gamma \big( J_\beta( v^2 ) \big) .
\end{align*}
The fact that $\Phi_t$ is well-defined follows from $ \beta < 1$, the bound
\begin{equation*}
\Big\Vert \big( \d \Gamma ( \< y \>^{\beta} ) + 1 \big)^{- 1} \d \Gamma \big( J_\beta ( v^2 ) \big) \big( \d \Gamma ( \< y \>^{\beta} ) + 1 \big)^{- 1} \Big\Vert \lesssim 1 ,
\end{equation*}
and Lemma \ref{a26}. We introduce the Heisenberg derivative
\begin{equation*}
D \Phi_t:= \partial_t \Phi_t - \ii \big[ \Phi_t , \widetilde{H} \big] ,
\end{equation*}
with the property $\partial_t \< \widetilde{u}_t, \Phi_t \widetilde{u}_t  \> =\< \widetilde{u}_t, D\Phi_t  \widetilde{u}_t  \>$.  We want to show that the leading term of $D\Phi_t$ is non-positive, while $\Phi_t\ge 0$. More precisely, we show below

\begin{lemma}\sl \label{a52}
Assume $0 \leq \beta < \delta < 1$, $0 \leq \gamma < \min ( ( 1 - 1 / \mathrm{c} ) \beta , 1 / 4 )$ and $0 < \varepsilon < 1 / 2 - 2 \gamma$. In the sense of quadratic forms on $\chi ( \widetilde{H} ) \D ( \d \Gamma ( \< y \>^{\beta} ) )$,
\begin{equation} \label{a53}
\Phi_t \geq t^{2 \gamma}  \d \Gamma ( F ( \vert v \vert ) ) ,
\end{equation}
and there exists $\mathrm{C}> 0$ such that
\begin{equation} \label{a54}
D \Phi_t \leq - \frac{\theta}{t} \Phi_t + \mathrm{C} t^{-1 - \delta + 2 \gamma} \d \Gamma ( \vert k \vert^{- \delta} ) + \mathrm{C} t^{- 1 - \varepsilon} ,
\end{equation}
where $\theta : = 2 ( ( 1 - 1 / \mathrm{c} ) \beta - \gamma ) > 0$.
\end{lemma}

Rewriting inequality \eqref{a54} in terms of quadratic forms on the vectors $\widetilde{u}_t = e^{-\ii t \widetilde{H}} \chi( \widetilde{H}) u$ and using $\Phi_{t} \geq 0$ and $\< \widetilde{u}_t, D\Phi_t\widetilde{u}_t\>=\partial_t \< \widetilde{u}_t, \Phi_t\widetilde{u}_t \>$, we obtain
\begin{equation*}
\partial_t \< \widetilde{u}_t, \Phi_t \widetilde{u}_t \> \lesssim  t^{-1 - \delta + 2 \gamma} \big\< \widetilde{u}_{t} , \d \Gamma ( \vert k \vert^{-\delta} ) \widetilde{u}_{t} \big\> +  t^{-1-\varepsilon} \Vert u \Vert^{2} .
\end{equation*}
It then follows from Lemma \ref{a20} that
\begin{equation*}
\partial_t \< \widetilde{u}_t, \Phi_t \widetilde{u}_t\> \lesssim  t^{- \frac{3}{5} (1 + \delta ) + 2 \gamma} \big( \big\Vert \d \Gamma ( \vert k \vert^{- \delta} )^{\frac{1}{2}} u \big\Vert^2 + \Vert u \Vert^2 \big)  +  t^{-1-\varepsilon} \Vert u \Vert^{2} .
\end{equation*}
Assuming $3 \delta > 10 \gamma + 2$, this yields
\begin{align*}
\partial_t \< \widetilde{u}_t , \Phi_t \widetilde{u}_t\> \lesssim  t^{- 1 - \widetilde{\varepsilon}} \big( \big\Vert \d \Gamma ( \vert k \vert^{- \delta} )^{\frac{1}{2}} u \big\Vert^2 + \Vert u \Vert^2 \big) ,
\end{align*}
for some $\widetilde{\varepsilon} > 0$. Integrating this inequality from $1$ to $t$, this implies
\begin{align*}
\< \widetilde{u}_t, \Phi_t \widetilde{u}_t \> \leq \big\< \widetilde{u}_{t = 1} , \Phi_{t = 1} \widetilde{u}_{t = 1} \big\> + \mathrm{C} \big( \big\Vert \d \Gamma ( \vert k \vert^{- \delta} )^{\frac{1}{2}} u \big\Vert^2 + \Vert u \Vert^2 \big) .
\end{align*}
Combined with \eqref{a53} and the fact
\begin{align*}
\Phi_{t=1} := \d \Gamma \Big( \Big( \frac{\vert y \vert}{\mathrm{c}} \Big)^{2 \beta} F \Big( \frac{\vert y \vert}{\mathrm{c}} \Big) \Big) \lesssim \d \Gamma \big( \vert y \vert^{2\beta} \big) ,
\end{align*}
which follows from the definition of $\Phi_{t}$ and Lemma \ref{a21}, this gives the desired inequality \eqref{a55}. This completes the proof of Theorem \ref{t2}.
\end{proof}

\begin{proof}[Proof of Lemma \ref{a52}]
Estimate \eqref{a53} is straightforward. To prove \eqref{a54}, we start with computing $D \Phi_t$. The relations below are understood in the sense of quadratic forms on $\chi ( \widetilde{H} ) \D ( \d \Gamma ( \< y \>^{\beta} ) )$. We compute
\begin{align}
D \Phi_t ={}& 2 t^{2 \gamma - 1} \ \d \Gamma \big( \gamma J_\beta ( v^2 ) - v^{2} J^{\prime}_\beta ( v^2 ) \big)  \label{a62}    \\
& - t^{2\gamma} \big[ \d \Gamma \big( J_\beta ( v^2 ) \big) , \ii \d \Gamma( \vert k \vert ) \big] \label{a63}      \\
& - t^{2\gamma} \Big[ \d \Gamma \big( J_\beta ( v^2 ) \big) , \ii \big( p + \widetilde{A}(x) \big)^2 + \ii E(x) \Big] .     \label{a64}
\end{align}

Consider the term given by \eqref{a63}. We have
\begin{equation*}
\big[ \d \Gamma \big( J_\beta ( v^2 ) \big) , \ii \d \Gamma( \vert k \vert ) \big] = \d \Gamma \big( \big[  J_\beta ( v^2  ) , \ii \vert k \vert \big] \big ),
\end{equation*}
and it follows from Lemma \ref{a65} that
\begin{equation} \label{a45}
\big[  J_\beta ( v^2  ) , \ii \vert k \vert \big] = \frac{1}{\mathrm{c} t} ( J^{\prime}_{\beta} )^{\frac{1}{2}} ( v^2 ) \big( v \cdot \widehat{k} + \widehat{k} \cdot v \big) ( J^{\prime}_{\beta} )^{\frac{1}{2}} ( v^2 ) + \CR ,
\end{equation}
where
\begin{equation} \label{a46}
\big\Vert \vert k \vert^{\frac{\delta}{2}} \CR \vert k \vert^{\frac{\delta}{2}} \big\Vert \lesssim t^{- 1 - \delta},
\end{equation}
for all $\beta < \delta \leq 1$. Observe that for all $w \in \D ( \vert v \vert^{\beta} ) = \D( \vert y \vert^{\beta} )$,
\begin{align*}
- \Big\< w , ( J^{\prime}_{\beta} )^{\frac{1}{2}} ( v^2 ) \big( v \cdot \widehat{k} + \widehat{k} \cdot v \big) ( J^{\prime}_{\beta} )^{\frac{1}{2}} ( v^2 ) w \Big\> & \leq 2 \Big\Vert \vert v \vert ( J^{\prime}_{\beta} )^{\frac{1}{2}} ( v^2 ) w \Big\Vert \Big\Vert ( J^{\prime}_{\beta} )^{\frac{1}{2}} ( v^2 ) w \Big\Vert     \nonumber \\
&\leq 2 \Big\Vert \vert v \vert ( J^{\prime}_{\beta} )^{\frac{1}{2}} ( v^2 ) w \Big\Vert^2 ,
\end{align*}
since $\supp ( J^{\prime}_{\beta} ) \subset [1 , \infty )$. This gives
\begin{equation} \label{a47}
- \d \Gamma \Big( \frac{1}{\mathrm{c} t} ( J^{\prime}_{\beta} )^{\frac{1}{2}} ( v^2 ) \big( v \cdot \widehat{k} + \widehat{k} \cdot v \big) ( J^{\prime}_{\beta} )^{\frac{1}{2}} ( v^2 ) \Big) \leq \frac{2}{\mathrm{c} t} \d \Gamma \big( v^2 J^{\prime}_{\beta} ( v^2 ) \big) .
\end{equation}
Combining \eqref{a45} with \eqref{a46} and \eqref{a47}, we get
\begin{equation} \label{a48}
- \big[ \d \Gamma \big( J_\beta ( v^2 ) \big) , \ii \d \Gamma ( \vert k \vert ) \big] \leq \frac{2}{\mathrm{c} t} \d \Gamma \big( v^2 J^{\prime}_{\beta} ( v^2 ) \big) - \mathrm{C} t^{-1 - \delta} \d \Gamma ( \vert k \vert^{-\delta} ) .
\end{equation}

It remains to estimate the term \eqref{a64}. Using the relation $\ii [ \d \Gamma ( b ) , \Phi ( G ) ] = \Phi ( \ii b G )$, we  compute
\begin{align*}
\eqref{a64} ={}& t^{2 \gamma} \big( p + \widetilde{A}(x) \big) \cdot \Phi \big( \ii J_\beta ( v^2 ) \widetilde{g}_{x} \big)      \\
&+ t^{2 \gamma} \Phi \big( \ii J_\beta ( v^2 ) \widetilde{g}_{x} \big) \cdot \big( p + \widetilde{A} (x) \big) + \ii t^{2 \gamma} \Phi \big( \ii J_\beta ( v^2 ) e_{x} \big) ,
\end{align*}
and hence Corollary \ref{a11} and Theorem \ref{a23} imply
\begin{align}\label{a51}
\Vert \one_{\supp ( \chi )} ( \widetilde{H} ) \eqref{a64} \one_{\supp ( \chi )} ( \widetilde{H} ) \Vert \lesssim t^{-1 - \varepsilon} ,
\end{align}
for all $0 < \varepsilon < 1/2 - 2 \gamma$.

The estimates \eqref{a48} and \eqref{a51}, together with \eqref{a62}--\eqref{a64}, imply
\begin{equation*}
D \Phi_t \leq 2 t^{2 \gamma - 1} \d \Gamma \Big( \gamma J_\beta ( v^2 ) - v^{2} J^{\prime}_\beta ( v^2 ) + \frac{1}{\mathrm{c}} v^2 J^{\prime}_{\beta} ( v^2) \Big) - \mathrm{C} t^{2 \gamma - 1  - \delta}  \d \Gamma ( \vert k \vert^{-\delta} ) - \mathrm{C} t^{- 1 - \varepsilon} ,
\end{equation*}
as a quadratic form on $\chi ( \widetilde{H} ) \D ( \d \Gamma ( \< y \>^{\frac{\beta}{2}} ) )$. Using
\begin{equation*}
v^{2} J^{\prime}_{\beta} ( v^{2} ) = \beta J_{\beta} ( v^{2} ) + \frac{1}{2} \vert v \vert^{2 \beta + 1} F^{\prime} ( \vert v \vert ) \geq \beta J_{\beta} ( v^{2} ) ,
\end{equation*}
this becomes
\begin{equation}
D \Phi_t \leq - \theta t^{- 1} \Phi_t - \mathrm{C} t^{2 \gamma - 1  - \delta}  \d \Gamma ( \vert k \vert^{-\delta} ) - \mathrm{C} t^{- 1 - \varepsilon} ,
\end{equation}
which concludes the proof of the lemma.
\end{proof}

\section{Estimates on interaction} \label{s3}

In this section we prove estimates on the interaction used, in particular, to prove \eqref{a51}. Recall that $\kappa \in \mathrm{C}_0^\infty( \mathbb{R}^3 )$ is the ultraviolet cut-off entering \eqref{a73} and the cut-off operator $F ( \vert v \vert  )$ is defined in \eqref{F}.

\begin{lemma}\sl \label{a3}
Let $a \in [ 0 , 3/2 )$, $b \in \mathbb{R}$, $c\geq 0$, $\kappa \in \mathrm{C}^{\infty}_{0} ( \R^{3} )$ and $\rho^{b}_{x} (k)$ be such that, for all $m \in \mathbb{N}^3$, $\vert \partial_{k}^m \rho^{b}_{x} ( k ) \vert \lesssim \vert k \vert^{b-|m|} \< x \>^{|m|}$. Assume that $b > a + c - 3/2$. Then, for all $d \in [ 0 , b - a - c + 3/2 )$,
\begin{equation*}
\forall x \in \R^{3}, \qquad \big\Vert \vert k \vert^{- a} |y|^c F ( \vert v \vert) \kappa (k) \rho^{b}_{x} (k) \big\Vert_{L^{2} ( \R^{3}_{k} )} \lesssim t^{- d} \< x \>^{a + c + d} .
\end{equation*}
\end{lemma}

\begin{proof}
Let $\ell_{x} ( k ) = \kappa (k) \rho^{b}_{x} ( k )$. Using Hardy's inequality (see Lemma \ref{a24}), we can write
\begin{align}
\big\Vert \vert k \vert^{- a} |y|^c F ( \vert v \vert ) \ell_{x} ( k ) \big\Vert
&\lesssim  \big\Vert \vert y \vert^{a+c} F ( \vert v \vert )  \ell_{x} ( k ) \big\Vert    \nonumber   \\
&\lesssim  \big\Vert F ( \vert v \vert ) \vert y \vert^{- d} \big\Vert_{L^{\infty}} \big\Vert \vert y \vert^{a + c + d}  \ell_{x} ( k ) \big\Vert    \nonumber   \\
&\lesssim t^{- d}  \big\Vert \vert y \vert^{a + c + d}  \ell_{x} ( k ) \big\Vert .   \label{a4}
\end{align}
Next, to handle fractional derivatives $\vert y \vert^{s}$, we use a dyadic decomposition of $\kappa$. Let $\varphi \in \mathrm{C}^{\infty}_{0} ( \R^{3} \setminus \{ 0 \} )$ be such that
\begin{equation}  \label{diad}
\forall k \in \supp ( \kappa ), \qquad \sum_{\nu \geq 1 \text{ dyadic}} \varphi ( \nu k ) = 1.
\end{equation}
For $n \in \N$, we have
\begin{equation*}
\big\Vert \vert y \vert^{n} \varphi ( \nu k ) \ell_{x} ( k ) \big\Vert \lesssim \sum_{i_{1} , \dots , i_{n} \in \{ 1 , 2 , 3 \}} \big\Vert y_{i_{1}} \cdots y_{i_{n}} \varphi ( \nu k ) \ell_{x} ( k ) \big\Vert ,
\end{equation*}
and $y_{i_{1}} \cdots y_{i_{n}} \varphi ( \nu k ) \ell_{x} ( k )$ can be written as a finite sum of terms of the form
\begin{equation*}
w = \nu^{\alpha} \widetilde{\varphi} ( \nu k ) \widetilde{\kappa} (k) \widetilde{\rho}^{b - \beta}_{x} ( k ) \< x \>^\beta ,
\end{equation*}
where $\alpha , \beta \in \N$ with $\alpha + \beta \leq n$, $\widetilde{\varphi} \in \mathrm{C}^{\infty}_{0} ( \R^{3} \setminus \{ 0 \})$, $\widetilde{\kappa} \in \mathrm{C}^{\infty}_{0} ( \R^{3} )$ and $\widetilde{\rho}^{b - \beta}_{x}$ is such that $\vert \widetilde{\rho}^{b - \beta}_{x} ( k ) \vert \lesssim \vert k \vert^{b - \beta} $. Then,
\begin{equation*}
\Vert w \Vert \lesssim \nu^{\alpha + \beta - b} \< x \>^{\beta} \Vert \widetilde{\varphi} ( \nu k ) \Vert \lesssim \nu^{\alpha + \beta - b - \frac{3}{2}} \< x \>^{\beta} \leq \nu^{n - b - \frac{3}{2}} \< x \>^{n} .
\end{equation*}
This gives
\begin{equation*}
\big\Vert \vert y \vert^{n} \varphi ( \nu k ) \ell_{x} ( k ) \big\Vert \lesssim \nu^{n - b - \frac{3}{2}} \< x \>^{n} .
\end{equation*}
Now, an interpolation argument implies that, for all $s \geq 0$,
\begin{equation} \label{a5}
\big\Vert \vert y \vert^{s} \varphi ( \nu k ) \ell_{x} ( k ) \big\Vert \lesssim \nu^{s - b - \frac{3}{2}} \< x \>^{s} .
\end{equation}
Combining \eqref{diad} and \eqref{a5}, we obtain
\begin{align}
\big\Vert  |y|^s \ell_{x} ( k ) \big\Vert
&\le  \sum_{\nu \geq 1 \text{ dyadic}} \big\Vert  |y|^s \varphi ( \nu k ) \ell_{x} ( k ) \big\Vert \nonumber \\
 &\lesssim  \sum_{\nu \geq 1 \text{ dyadic}} \nu^{s - b - \frac{3}{2}} \< x \>^{s} \lesssim  \< x \>^{s},   \label{a6}
\end{align}
provided that $b + 3/2 - s > 0$. Taking $s=a+c+d$ and $d \in [ 0 , b - a - c + 3/2 )$ and recalling \eqref{a4}, we arrive at the statement of the lemma.
\end{proof}

Recall that the coupling functions $q_{x}$, $\widetilde{g}_{x}$ and $e_{x}$ are defined at the beginning of Section \ref{s2} and satisfy \eqref{a56}--\eqref{a58}.

\begin{corollary}\sl \label{a11}
For all $0 < \mu < 1/2$, $0 \leq \beta \leq 1/2$ and $\varepsilon > 0$,
\begin{align}
&\Big\Vert \Phi \big( \ii F  (  \vert v \vert ) q_{x} \big) \< x \>^{-\tau_1} ( H_f + 1 )^{-\frac{1}{2}} \Big\Vert \lesssim t^{-d}, \quad 0 \leq d < \frac{1}{2} ,      \label{a8} \\
&\Big\Vert \Phi \big( \ii \vert y \vert^{2\beta} F  ( \vert v \vert ) \widetilde{g}_{x} \big) \< x \>^{-\tau_2} ( H_f + 1 )^{-\frac{1}{2}} \Big\Vert \lesssim t^{-d}, \quad 0 \leq d < \frac{3}{2} - 2 \beta ,        \label{a9} \\
&\Big\Vert \Phi \big( \ii \vert y \vert^{2\beta} F  ( \vert v \vert ) e_{x} \big) \< x \>^{-\tau_3} ( H_f + 1 )^{-\frac{1}{2}} \Big\Vert \lesssim t^{-d}, \quad 0 \leq d < \frac{3}{2} - 2 \beta ,    \label{a10}
\end{align}
where $\tau_1 = 3/2 + d$, $\tau_2= 1/2 + \mu^{-1} + 2 \beta + d$ and $\tau_3 = 3/2 + 2\beta + d$.
\end{corollary}

\begin{proof}
It follows from Lemma \ref{a22} that, for all $u \in \mathcal{H} = \mathrm{L}^2( \mathbb{R}^3 ; \mathcal{F} )$,
\begin{align}
\Big\Vert \Phi \big( \ii F ( \vert v \vert ) q_{x} \big) & \< x \>^{- \tau_1} ( H_f + 1 )^{-\frac{1}{2}} u \Big\Vert^2     \notag \\
&\lesssim \int_{\mathbb{R}^3}  \< x \>^{-2\tau_1} \Big( \big\Vert \vert k \vert^{-\frac{1}{2}} F ( \vert v \vert ) q_{x} ( k,\lambda) \big\Vert^2_{ \mathrm{L}^2( \mathbb{R}^3 ; \C^{2} ) }    \notag \\
&\qquad \qquad \qquad + \big\Vert F  ( \vert v \vert ) q_{x} (k,\lambda) \big\Vert^2_{ \mathrm{L}^2( \mathbb{R}^3 ; \C^{2} ) } \Big) \Vert u(x) \Vert^2_{ \mathcal{F} } \, \d x .
\end{align}
Using \eqref{a56} and applying Lemma \ref{a3} with $a = 1/2$, $b = - 1 / 2$, $c = 0$ to the first term on the right hand side, and with $a = 0$, $b = - 1 / 2$, $c =0 $ to the second term, we obtain
\begin{align}
\Big\Vert \Phi \big( \ii F ( & \vert v \vert ) q_{x} \big) \< x \>^{- \tau_1} ( H_f + 1 )^{-\frac{1}{2}} u \Big\Vert^2 \lesssim t^{-2d} \int_{ \mathbb{R}^3 } \Vert u(x) \Vert^2_{ \mathcal{F} } \, \d x = t^{-2d} \Vert u \Vert^2 ,
\end{align}
which gives \eqref{a8}. To prove \eqref{a9} or \eqref{a10}, we proceed as above, applying Lemma \ref{a3} with $a = 1 / 2$, $b = 1 / 2$, $c = 2 \beta$ and with  $a = 0$, $b = 1 / 2$, $c = 2 \beta$.
\end{proof}

\section{Control of small momenta} \label{s4}

In this section we estimate the growth of $\d \Gamma ( \vert k \vert^{- \delta} )$ (for $-1 < \delta < 3/2$) along the evolution, which was used in the proof of Theorem \ref{t2}. The proof of the following lemma is similar to \cite[(4.8)]{Ge02_01}.

\begin{lemma}\sl \label{a20}
Let $- 1 < \delta <  3 / 2$ and $\chi \in \mathrm{C}_0^\infty ( ( - \infty , \Sigma ) )$. Then, for all $u \in \D ( \d \Gamma ( \vert k \vert^{- \delta} )^{\frac{1}{2}} )$,
\begin{equation*}
\Big\< e^{-\ii t \widetilde{H}} \chi( \widetilde{H} ) u , \d \Gamma ( \vert k \vert^{- \delta} ) e^{-\ii t \widetilde{H}} \chi( \widetilde{H} ) u \Big\> \lesssim t^{\frac{2 ( 1 + \delta )}{5}} \big( \big\Vert \d \Gamma ( \vert k \vert^{- \delta} )^{\frac{1}{2}} u \big\Vert^2 + \Vert u \Vert^2 \big) .
\end{equation*}
\end{lemma}

\begin{proof}
Let $h \in \mathrm{C}^\infty( [0,\infty) ; \mathbb{R} )$ be a decreasing function such that $h(s) = 1$ on $[ 0 , 1 ]$ and $h(s) = 0$ on $[ 2 , + \infty )$, and let $\bar h = \mathds{1} - h$. For $\nu>0$, we decompose
\begin{equation} \label{a17}
\d \Gamma ( \vert k \vert^{- \delta} ) = \d \Gamma \big( \vert k \vert^{- \delta} h( t^{\nu} \vert k \vert ) \big) +  \d \Gamma \big( \vert k \vert^{- \delta} \bar h (t^{\nu} \vert k \vert ) \big) .
\end{equation}
As before, let $\widetilde{u}_{t} = e^{-\ii t \widetilde{H}} \chi ( \widetilde{H} ) u$. The contribution of the second term of \eqref{a17} is estimated as
\begin{equation} \label{a18}
\big\< \widetilde{u}_{t} , \d \Gamma \big( \vert k \vert^{- \delta} \bar h (t^{\nu} \vert k \vert ) \big) \widetilde{u}_{t} \big\> \leq t^{( 1 + \delta ) \nu} \big\< \widetilde{u}_{t} , \d \Gamma \big( \vert k \vert \bar h ( t^{\nu} \vert k \vert ) \big) \widetilde{u}_{t} \big\> \lesssim t^{( 1 + \delta ) \nu} \Vert u \Vert^2 ,
\end{equation}
since $\d \Gamma ( \vert k \vert \bar h ( t^{\nu} \vert k \vert ) ) \chi ( \widetilde{H} )$ is bounded. To estimate the first term, we use the propagation observable
\begin{equation*}
\Psi_t : = \d \Gamma \big( \vert k \vert^{- \delta} h( t^{\nu} \vert k \vert ) \big) ,
\end{equation*}
and compute
\begin{equation*}
D \Psi_t = - \big[ \Psi_t , \ii \widetilde{H} \big] + \nu t^{\nu-1}  \d \Gamma \big( \vert k \vert^{1 - \delta} h^{\prime} ( t^{\nu} \vert k \vert ) \big) \leq - \big[ \Psi_t , \ii \widetilde{H} \big] ,
\end{equation*}
since $h^{\prime} \leq 0$. The commutator above can be expressed as follows
\begin{align*}
\big[ \Psi_t , \ii \widetilde{H} \big] ={}& \big( p + \widetilde{A} (x) \big) \cdot \Phi \big( \vert k \vert^{- \delta} h( t^{\nu} \vert k \vert ) \widetilde{g}_{x} \big)       \\
&+ \Phi \big( \ii \vert k \vert^{- \delta} h( t^{\nu} \vert k \vert ) \widetilde{g}_{x} \big) \cdot \big( p + \widetilde{A} (x) \big)        \\
&+ \ii \Phi \big( \ii \vert k \vert^{- \delta} h( t^{\nu} \vert k \vert ) e_{x} \big) .
\end{align*}
Using \eqref{a57} and Lemma \ref{a22}, we find that for all $x \in \mathbb{R}^3$,
\begin{align*}
\Big\Vert \Phi \big( \ii \vert k \vert^{- \delta} h( t^{\nu} \vert k \vert ) \widetilde{g}_{x} \big) ( H_f + 1 )^{-\frac{1}{2}} \Big\Vert \leq{}& \big\Vert \vert k \vert^{- \delta} h( t^{\nu} \vert k \vert ) \widetilde{g}_{x} ( k , \lambda ) \big\Vert_{\mathrm{L}^2 ( \mathbb{R}^3 ; \C^{2} )}     \\
&+ \big\Vert \vert k \vert^{- \delta - \frac{1}{2}} h( t^{\nu} \vert k \vert ) \widetilde{g}_{x} ( k , \lambda ) \big\Vert_{\mathrm{L}^2 ( \mathbb{R}^3 ; \C^{2} )}     \\
\lesssim{}& \big\Vert \vert k \vert^{- \delta} h( t^{\nu} \vert k \vert ) \kappa(k) \big\Vert_{\mathrm{L}^2 ( \mathbb{R}^3 ; \C^{2} )} \< x \>^{\frac{1}{\mu}}     \\
\lesssim{}& t^{- ( \frac{3}{2} - \delta ) \nu} \< x \>^{\frac{1}{\mu}} .
\end{align*}
Likewise, using \eqref{a58} and Lemma \ref{a22}, we obtain
\begin{align*}
\Big\Vert \Phi \big( \ii \vert k \vert^{- \delta} h( t^{\nu} \vert k \vert ) e_{x} \big) ( H_f + 1 )^{-\frac{1}{2}} \Big\Vert \leq{}& \big\Vert \vert k \vert^{- \delta} h( t^{\nu} \vert k \vert ) e_{x} ( k , \lambda ) \big\Vert_{\mathrm{L}^2 ( \mathbb{R}^3 ; \C^{2} )}  \\
&+ \big\Vert \vert k \vert^{- \delta - \frac{1}{2}} h( t^{\nu} \vert k \vert ) e_{x} ( k , \lambda ) \big\Vert_{\mathrm{L}^2 ( \mathbb{R}^3 ; \C^{2} )}  \\
\lesssim{}& t^{- ( \frac{3}{2} - \delta ) \nu} \< x \> .
\end{align*}
The last two estimates, Theorem \ref{a23}, $\Vert ( p + \widetilde{A} (x) ) \chi ( \widetilde{H} ) \Vert \lesssim 1$ and $\partial_{t} \< \widetilde{u}_{t} , \Psi_{t} \widetilde{u}_{t} \> = \< \widetilde{u}_{t} , D \Psi_{t} \widetilde{u}_{t} \>$ imply
\begin{equation*}
\partial_{t} \< \widetilde{u}_{t} , \Psi_{t} \widetilde{u}_{t} \> \lesssim t^{- ( \frac{3}{2} - \delta ) \nu} \Vert u \Vert^2 ,
\end{equation*}
and hence, assuming $( \frac{3}{2} - \delta ) \nu < 1$,
\begin{equation} \label{a19}
\< \widetilde{u}_{t} , \Psi_{t} \widetilde{u}_{t} \> \lesssim t^{- ( \frac{3}{2} - \delta ) \nu + 1} \Vert u \Vert^2 + \big\Vert \d \Gamma ( \vert k \vert^{- \delta} )^{\frac{1}{2}} u \big\Vert^2.
\end{equation}
The statement of the lemma follows from \eqref{a17}, \eqref{a18} and \eqref{a19}, after choosing $\nu = 2 / 5$.
\end{proof}

\section{Some commutators estimates} \label{s5}

In this part, we estimate some commutators appearing in Section \ref{s2}. As usual, for $\rho \in \mathbb{R}$, we define the set of functions
\begin{equation} \label{a44}
S^\rho( \mathbb{R} ) := \big\{ f \in \mathrm{C}^\infty( \mathbb{R} ) ; \ \big\vert \partial_s^n f(s) \big\vert \leq \mathrm{C}_n \< s \>^{\rho-n} \text{ for } n \geq 0 \big\} .
\end{equation}

\begin{lemma}\sl \label{a12}
Let $G \in \mathrm{S}^\rho( \mathbb{R} )$ with $\rho < 0$ and $\max ( 1 + 2 \rho , 0) < \delta \leq 1$. We have
\begin{equation*}
\big\Vert \vert k \vert^{\frac{\delta}{2}} \big[ G  ( v^2 ) ,  y \cdot \widehat{k} + \widehat{k} \cdot y \big] \vert k \vert^{\frac{\delta}{2}} \big\Vert \lesssim t^{1 - \delta} .
\end{equation*}
\end{lemma}

\begin{proof}
Let $\widetilde{G}$ denote an almost analytic extension of $G$ such that $\widetilde{G} \vert_{\mathbb{R}} = G$,
\begin{equation} \label{a66}
\supp \widetilde{G} \subset \big\{ z \in \C ; \ \vert \im z \vert \leq \mathrm{C} \< \re z \> \big\} ,
\end{equation}
$\vert \widetilde{G} (z) \vert \leq \mathrm{C} \< \re z \>^{\rho}$ and, for all $n \in \N$,
\begin{equation} \label{a67}
\Big\vert \frac{\partial \widetilde{G}}{\partial \bar z} (z) \Big\vert \leq \mathrm{C}_n \< \re z \>^{\rho - 1 - n} \vert \im z \vert^n .
\end{equation}
Using the Helffer--Sj{\"o}strand formula (see e.g. \cite{DeGe97_01,HuSi00_01})
\begin{equation*}
G ( v^{2} ) = \frac{1}{\pi} \int \frac{\partial \widetilde{G} (z)}{\partial \bar z} ( v^2 - z )^{-1} \ddre z \ddim z ,
\end{equation*}
we can write
\begin{align}
\big[ G ( v^2 ) , y \cdot \widehat{k} & + \widehat{k} \cdot y \big]   \nonumber \\
&= \frac{1}{\pi} \int \frac{\partial \widetilde{G} (z)}{\partial \bar z} \big[ ( v^2 - z )^{-1} , y \cdot \widehat{k} + \widehat{k} \cdot y \big] \ddre z \ddim z \notag   \\
&= - \frac{1}{\pi \mathrm{c}^{2} t^2} \int \frac{\partial \widetilde{G} (z)}{\partial \bar z} ( v^2 - z )^{-1} \big[ y^2 , y \cdot \widehat{k} + \widehat{k} \cdot y \big] ( v^2 - z )^{-1} \ddre z \ddim z . \label{a1}
\end{align}
Let us first prove that
\begin{equation} \label{a13}
( v^2 - z )^{-1} \big[ y^2 , y \cdot \widehat{k} + \widehat{k} \cdot y \big] ( v^2 - z )^{-1} \vert k \vert = \CO \big( t^{2} \vert z \vert^{2} \vert \im z \vert^{-3} \big) .
\end{equation}
A direct calculation gives
\begin{align}
\frac{1}{\ii} \big[ y^2 , y \cdot \widehat{k} + \widehat{k} \cdot y \big] ={}& y^2 \vert k \vert^{-1} + \vert k \vert^{-1} y^2 + 2 \sum_i y_i \vert k \vert^{-1} y_i    \nonumber  \\
&- \sum_{i,j} y_i y_j \big( k_i k_j \vert k \vert^{-3} \big) + 2 y_i \big( k_i k_j \vert k \vert^{-3} \big) y_j + \big( k_i k_j \vert k \vert^{-3} \big) y_i y_j . \label{a2}
\end{align}
Using Hardy's inequality (see Lemma \ref{a24}) and the functional calculus, we get
\begin{align}
( v^2  - z )^{-1} \vert k \vert ={}& \vert k \vert ( v^2 - z )^{-1}    \nonumber   \\
&- \frac{\ii}{t^{2}} ( v^2 - z )^{-1} \big( 2\widehat{k} \cdot y + 2 \ii \vert k \vert^{-1} \big) ( v^2 - z )^{-1}     \nonumber  \\
={}& \vert k \vert \CO \big( \vert z \vert \vert \im z \vert^{-2} \big) ,   \label{a68}  \\
y_{i} ( v^2 - z )^{-1} \vert k \vert ={}& \vert k \vert y_{i} ( v^2 - z )^{-1} + \ii \widehat{k}_{i} ( v^2 - z )^{-1}    \nonumber  \\
&- \frac{\ii}{t^{2}} y_{i} ( v^2 - z )^{-1} \big( 2\widehat{k} \cdot y + 2 \ii \vert k \vert^{-1} \big) ( v^2 - z )^{-1}    \nonumber  \\
={}& \vert k \vert \CO \big( t \vert z \vert^{\frac{3}{2}} \vert \im z \vert^{- 2} \big) ,    \nonumber  \\
y_{i} y_{j} ( v^2  - z )^{-1} \vert k \vert ={}& \vert k \vert y_{i} y_{j} ( v^2 - z )^{-1}     \nonumber  \\
&+ \big( \ii \widehat{k}_{i} y_{j} + \ii \widehat{k}_{j} y_{i} + k_{i} k_{j} \vert k \vert^{-3} - \delta_{i,j} \vert k \vert^{-1} \big) ( v^2 - z )^{-1}    \nonumber  \\
&- \frac{\ii}{t^{2}} y_{i} y_{j} ( v^2 - z )^{-1} \big( 2\widehat{k} \cdot y + 2 \ii \vert k \vert^{-1} \big) ( v^2 - z )^{-1}     \nonumber  \\
={}& \vert k \vert \mathcal{O} \big( t^{2} \vert z \vert \vert \im z \vert^{-1} \big) + \CO \big( t \vert z \vert^{\frac{3}{2}} \vert \im z \vert^{-2} \big) ,  \nonumber
\end{align}
and
\begin{gather*}
( v^2 - z )^{-1} y_{i} y_{j} = \CO \big( t^{2} \vert z \vert \vert \im z \vert^{-1} \big) ,  \qquad    ( v^2 - z )^{-1} y_{i} = \CO \big( t \vert z \vert^{\frac{1}{2}} \vert \im z \vert^{-1} \big) ,    \\
( v^2 - z )^{-1} \vert k \vert^{-1} = \CO \big( t \vert z \vert^{\frac{1}{2}} \vert \im z \vert^{-1} \big) ,     \qquad   ( v^2 - z )^{-1} = \CO \big( \vert \im z \vert^{-1} \big) .
\end{gather*}
Combining \eqref{a2} with the previous estimates, we obtain \eqref{a13}.

Now, using again \eqref{a2} and the previous estimates, one easily verifies that
\begin{equation} \label{a14}
( v^2 - z )^{-1} \big[ y^2 , y \cdot \widehat{k} + \widehat{k} \cdot y \big] ( v^2 - z )^{-1} = \CO \big( t^{3} \vert z \vert^{\frac{3}{2}} \vert \im z \vert^{-2} \big) .
\end{equation}
By an interpolation argument, we then obtain from \eqref{a13} (and its adjoint) and \eqref{a14} that
\begin{equation} \label{a15}
\vert k \vert^{\frac{\delta}{2}} ( v^2 - z )^{-1} \big[ y^2 , y \cdot \widehat{k} + \widehat{k} \cdot y \big] ( v^2 - z )^{-1} \vert k \vert^{\frac{\delta}{2}} = \CO \big( t^{3 - \delta} \vert z \vert^{\frac{3}{2} + \frac{\delta}{2}} \vert \im z \vert^{- 2 - \delta} \big) ,
\end{equation}
for all $0 \leq \delta \leq 1$.

Introducing \eqref{a15} into \eqref{a1} gives
\begin{align}
\big\Vert \vert k \vert^{\frac{\delta}{2}} \big[ G ( v^2 ) , y \cdot \widehat{k} + \widehat{k} \cdot y \big] \vert k \vert^{\frac{\delta}{2}} u \big\Vert &\lesssim t^{1 - \delta} \int \Big\vert \frac{\partial \widetilde{G} (z)}{\partial \bar z} \Big\vert \vert z \vert^{\frac{3}{2} + \frac{\delta}{2}} \vert \im z \vert^{- 2 - \delta} \Vert u \Vert \, \ddre z \, \ddim z     \notag \\
&\lesssim t^{1 - \delta} \int \< \re z \>^{- \frac{1}{2} + \rho - \frac{\delta}{2}} \Vert u \Vert \, \ddre z \lesssim t^{1 - \delta} \Vert u \Vert ,
\end{align}
provided that $\delta > 1 + 2 \rho$, which concludes the proof of the lemma.
\end{proof}

\begin{lemma}\sl \label{a65}
Let $G \in \mathrm{S}^\rho( \mathbb{R} )$ with $\rho < 1$ and $\max ( 1 + 2 \rho , 0) < \delta \leq 1$. We have
\begin{equation*}
\big[ G (  v^2  ) , \ii \vert k \vert \big] = \frac{1}{ \mathrm{c} t } G^{\prime}  ( v^2 ) \big( v \cdot \widehat{k} + \widehat{k} \cdot v \big) + \CR ,
\end{equation*}
as a quadratic form on $\D ( \vert y \vert^{2 \rho} ) \cap D ( \vert k \vert )$, with
\begin{equation*}
\big\Vert \vert k \vert^{\frac{\delta}{2}} \CR \vert k \vert^{\frac{\delta}{2}} \big\Vert \lesssim t^{- \delta} .
\end{equation*}
\end{lemma}

\begin{proof}
Since $\rho$ may be non-negative, we cannot directly express $G ( v^{2} )$ with the Helffer--Sj{\"o}strand formula. Therefore, we use an artificial cut-off. Consider $\varphi \in \mathrm{C}_{0}^{\infty} ( \R ; [ 0 , 1 ] )$ equal to $1$ near $0$ and $\varphi_{R} ( \cdot ) = \varphi( \cdot / R )$ for $R > 0$. Let $\widetilde{G}$ (resp. $\widetilde{\varphi} \in \mathrm{C}_{0}^{\infty} ( \C )$) be an almost analytic extension of $G$ (resp. $\varphi$) as in \eqref{a66}--\eqref{a67}. Then, as a quadratic form on $\D ( \vert y \vert^{2 \rho} ) \cap D ( \vert k \vert )$,
\begin{equation}
\big[ G (  v^2  ) , \ii \vert k \vert \big] = \slim_{R \to \infty} \big[ ( \varphi_{R} G ) (  v^2  ) , \ii \vert k \vert \big] ,
\end{equation}
where
\begin{align}
\big[ ( \varphi_{R} G ) (  v^2  ) , \ii \vert k \vert \big] &= \frac{1}{\pi} \int \frac{\partial ( \widetilde{\varphi}_{R} \widetilde{G} ) (z)}{\partial \bar z} \big[ (  v^2 - z )^{-1} , \ii \vert k \vert \big] \ddre z \ddim z    \nonumber \\
&= - \frac{1}{\pi} \int \frac{\partial ( \widetilde{\varphi}_{R} \widetilde{G} ) (z)}{\partial \bar z} (  v^2 - z )^{-1} \big[ v^{2} , \ii \vert k \vert \big] (  v^2 - z )^{-1} \ddre z \ddim z     \nonumber \\
&= \frac{1}{\pi \mathrm{c} t} \int \frac{\partial ( \widetilde{\varphi}_{R} \widetilde{G} ) (z)}{\partial \bar z} (  v^2 - z )^{-1} \big( v \cdot \widehat{k} + \widehat{k} \cdot v \big) (  v^2 - z )^{-1} \ddre z \ddim z    \nonumber \\
&= \frac{1}{\mathrm{c} t} ( \varphi_{R} G )^{\prime} ( v^{2} ) \big( v \cdot \widehat{k} + \widehat{k} \cdot v \big) + \CR_{R} ,  \label{a72}
\end{align}
and
\begin{align}
\CR_{R} &= \frac{1}{\pi \mathrm{c} t} \int \frac{\partial ( \widetilde{\varphi}_{R} \widetilde{G} ) (z)}{\partial \bar z} (  v^2 - z )^{-1} \big[ v \cdot \widehat{k} + \widehat{k} \cdot v , (  v^2 - z )^{-1} \big] \ddre z \ddim z    \nonumber \\
&= \frac{1}{\pi \mathrm{c}^{4} t^{4}} \int \frac{\partial ( \widetilde{\varphi}_{R} \widetilde{G} ) (z)}{\partial \bar z} (  v^2 - z )^{-2} \big[ y^{2} ,  v \cdot \widehat{k} + \widehat{k} \cdot v \big] (  v^2 - z )^{-1} \ddre z \ddim z . \label{a69}
\end{align}

From \eqref{a13}, \eqref{a68} and \eqref{a14}, we obtain
\begin{gather*}
( v^2 - z )^{-2} \big[ y^2 , y \cdot \widehat{k} + \widehat{k} \cdot y \big] ( v^2 - z )^{-1} \vert k \vert = \CO \big( t^{2} \vert z \vert^{2} \vert \im z \vert^{- 4} \big) ,  \\
\vert k \vert ( v^2 - z )^{-2} \big[ y^2 , y \cdot \widehat{k} + \widehat{k} \cdot y \big] ( v^2 - z )^{-1} = \CO \big( t^{2} \vert z \vert^{3} \vert \im z \vert^{- 5} \big) ,   \\
( v^2 - z )^{-2} \big[ y^2 , y \cdot \widehat{k} + \widehat{k} \cdot y \big] ( v^2 - z )^{-1} = \CO \big( t^{3} \vert z \vert^{\frac{3}{2}} \vert \im z \vert^{-3} \big) .
\end{gather*}
Then, an interpolation argument gives
\begin{equation} \label{a70}
\vert k \vert^{\frac{\delta}{2}} ( v^2 - z )^{-2} \big[ y^2 , y \cdot \widehat{k} + \widehat{k} \cdot y \big] ( v^2 - z )^{-1} \vert k \vert^{\frac{\delta}{2}} = \CO \big( t^{3 - \delta} \vert z \vert^{\frac{3}{2} ( 1 + \delta )} \vert \im z \vert^{- 3 - 2 \delta} \big) .
\end{equation}
On the other hand, for all $n \in \N$,
\begin{equation} \label{a71}
\Big\vert \frac{\partial ( \widetilde{\varphi}_{R} \widetilde{G} )}{\partial \bar z} (z) \Big\vert \leq \mathrm{C}_n \< \re z \>^{\rho - 1 - n} \vert \im z \vert^n ,
\end{equation}
where $\mathrm{C}_n > 0$ does not depend on $R \geq 1$. Using \eqref{a69} together with \eqref{a70} and \eqref{a71}, there exists $\mathrm{C} > 0$ such that
\begin{equation*}
\big\Vert \vert k \vert^{\frac{\delta}{2}} \CR_{R} \vert k \vert^{\frac{\delta}{2}} \big\Vert \leq \mathrm{C} t^{- 1 - \delta} ,
\end{equation*}
for all $R \geq 1$. Eventually, since $( \varphi_{R} G )^{\prime} ( v^{2} )$ converges strongly to $G^{\prime} ( v^{2} )$ on $\D ( \vert v \vert^{2 \rho} )$, the lemma follows from \eqref{a72} and the previous estimate.
\end{proof}

\appendix

\section{Standard estimates} \label{s6}

In this part, we state the following well-known results. Recall that $y = \ii \nabla_{k}$.

\begin{lemma}[Hardy's inequality on $\R^{3}$]\sl \label{a24}
For all $0 \leq s < 3/2$, we have $\D ( \vert y \vert^{s} ) \subset \D ( \vert k \vert^{-s} )$ and, for all $u \in \D ( \vert y \vert^{s} )$,
\begin{equation*}
\big\Vert \vert k \vert^{-s} u \big\Vert \lesssim \big\Vert \vert y \vert^{s} u \big\Vert .
\end{equation*}
\end{lemma}

\begin{lemma}\sl \label{a21}
Let $a,b$ be two self-adjoint operators on $\mathrm{L}^2( \mathbb{R}^3 ; \C^{2} )$ with $b \geq 0$, $\D (b) \subset \D (a)$ and $\Vert a \varphi \Vert \leq \Vert b \varphi \Vert$ for all $\varphi \in \D (b)$. Then $\D ( \d\Gamma(b) ) \subset \D( \d\Gamma(a) )$ and $\Vert \d \Gamma(a) \Phi \Vert \leq \Vert \d \Gamma(b) \Phi \Vert$ for all $\Phi \in \D ( \d\Gamma(b) )$.
\end{lemma}

\begin{lemma}\sl \label{a7}
For any $f \in \mathrm{L}^2( \mathbb{R}^3 ; \C^{2} )$, the operators $a(f) ( N +1 )^{-1/2}$ and $a^{*} (f) (N+1 )^{-1/2}$ extend to bounded operators on $\mathcal{H}$ satisfying
\begin{gather*}
\big\Vert a(f) (	N+1 )^{- \frac{1}{2} } \big\Vert \leq \Vert f \Vert ,    \\
\big\Vert a^{*} (f) ( N+1 )^{- \frac{1}{2} } \big\Vert \leq \sqrt{2} \Vert f \Vert .
\end{gather*}
\end{lemma}

\begin{lemma}\sl \label{a22}
Let $f \in \mathrm{L}^2( \mathbb{R}^3 ; \C^{2} )$ be such that $(k,\lambda) \mapsto \vert k \vert^{-1/2} f(k,\lambda) \in \mathrm{L}^2( \mathbb{R}^3 ; \C^{2} )$. Then, the operators $a(f) ( H_f + 1 )^{-1/2}$ and $a^{*} (f) ( H_f + 1 )^{-1/2}$ extend to bounded operators on $\mathcal{H}$ satisfying
\begin{gather*}
\big\Vert a(f) (	H_f + 1)^{- \frac{1}{2} } \big\Vert \leq \big\Vert \vert k \vert^{-\frac{1}{2}} f \big\Vert ,     \\
\big\Vert a^{*} (f) (H_f + 1 )^{- \frac{1}{2} } \big\Vert \leq \big\Vert \vert k \vert^{-\frac{1}{2}} f \big\Vert + \Vert f \Vert .
\end{gather*}
\end{lemma}

\section{Properties of the Hamiltonians $H$ and $\widetilde{H}$} \label{s7}

In this appendix, we collect a few properties of the Hamiltonians $H$ and $\widetilde{H}$. We begin with the following two important results.

\begin{theorem}[Self-adjointness \cite{HaHe08_01,Hi02_01}]\sl \label{a25}
The Hamiltonians $H$ and $\widetilde{H}$ are self-adjoint operators on the domain
\begin{equation*}
\D ( H ) = \D ( \widetilde{H} ) = \D \big( p^2 +  H_f \big) .
\end{equation*}
\end{theorem}

The fact that $H$ is self-adjoint on $\D(p^2+H_f)$ is proved in \cite{Hi02_01} by functional integral methods. Another proof is given in \cite{HaHe08_01} using abstract results based on commutator arguments. Self-adjointness of $\widetilde{H}$ on $\D( p^2 + H_f )$ is another application of \cite{HaHe08_01}, using that $\vert \widetilde{g}_{x} ( k , \lambda ) \vert \lesssim \kappa ( k ) \vert k \vert^{- \frac{1}{2} - \mu}$ with $0 < \mu < 1/2$.

\begin{theorem}[Exponential decay below the ionization threshold \cite{Gr04_01}]\sl \label{a23}
For all real numbers $\delta$ and $\xi$ such that $\xi + \delta^2 < \Sigma$,
\begin{equation*}
\big\Vert e^{\delta \vert x \vert} \mathds{1}_{(-\infty , \xi ]} ( H ) \big\Vert = \big\Vert e^{\delta \vert x \vert} \mathds{1}_{(-\infty , \xi ]} ( \widetilde{H} ) \big\Vert \lesssim 1 .
\end{equation*}
\end{theorem}

We now establish a property used in the proof of Theorem \ref{t1}. It shows in particular that the propagation observable $\Phi_t$ of Theorem \ref{t1} is well-defined. Since $H$ and $\widetilde{H}$ are \textit{not} of class $\mathrm{C}^1( \d \Gamma ( \< y \> ) )$, the proof of the next proposition is not straightforward. We refer to \cite{AmBoGe96_01} for the definition of the class $\mathrm{C}^1( \cdot )$ and its properties.

\begin{proposition}\sl \label{a26}
Let $H^\#$ denote either $H$ or $\widetilde{H}$. For all $\chi \in \mathrm{C}_0^\infty( (- \infty , \Sigma ) )$ and $0 \leq \beta < 1$, we have
\begin{equation*}
\chi ( H^{\#} ) \mathcal{D} \big( \d \Gamma ( \< y \>^{\beta} ) \big) \subset \mathcal{D} \big( \d \Gamma ( \< y \>^{\beta} ) \big) .
\end{equation*}
\end{proposition}

\begin{remark}\sl
The allowed power of $\< y \>$ in Proposition \ref{a26} is related to the infrared singularity of the interaction. More precisely, the requirement that $\beta < 1$ is due to the fact that the infrared behavior of the interaction in $H$ is of order $\vert k \vert^{-1/2}$. On the other hand, since the infrared behavior of the interaction in $\widetilde{H}$ is of order $\vert k \vert^{1 / 2}$, one could in fact show that
\begin{equation*}
\chi ( \widetilde{H} ) \D \big( \d \Gamma ( \< y \>^{\beta} ) \big) \subset \D \big( \d \Gamma ( \< y \>^{\beta} ) \big) ,
\end{equation*}
for any $0 \leq \beta < 2$. For our purpose, however, the stated result is sufficient.
\end{remark}

We shall need the following two lemmas to prove Proposition \ref{a26}.

\begin{lemma}\sl \label{a29}
Let $H^\#$ denote either $H$ or $\widetilde{H}$. Then
\begin{equation*}
H^\# \in \mathrm{C}^1(N) .
\end{equation*}
In particular, for all $\chi \in \mathrm{C}_0^\infty( \R )$,
\begin{equation*}
\chi ( H^\# ) \mathcal{D} ( N ) \subset \mathcal{D} ( N ).
\end{equation*}
\end{lemma}

\begin{proof}
Let us prove that $\widetilde{H} \in \mathrm{C}^1(N)$. Since $\D(\widetilde{H}) = \D ( p^2 + H_f )$ and since $N$ commutes with $p^2 + H_f$, we obviously have that
\begin{equation*}
\forall s \in \mathbb{R}, \qquad e^{\ii s N} \D( \widetilde{H} ) \subset \D ( \widetilde{H} ).
\end{equation*}
Therefore, by \cite[Theorem 6.3.4]{AmBoGe96_01} (see also \cite{GeGe99_01}), it suffices to prove that
\begin{equation} \label{a30}
\big\vert \big\< \widetilde{H} u , N u \big\> - \big\< N u , \widetilde{H} u \big\> \big\vert \lesssim \big( \big\Vert \widetilde{H} u \big\Vert^2 + \Vert u \Vert^2 \big) ,
\end{equation}
for all $u \in \D ( \widetilde{H} ) \cap \D ( N )$. In the sense of quadratic forms on $\D ( \widetilde{H} ) \cap \D ( N )$, we can compute
\begin{equation*}
\big[ \widetilde{H} , N \big] = \ii \big( p + \widetilde{A}(x) \big) \cdot \Phi ( \ii \widetilde{g}_{x} ) + \ii \Phi ( \ii \widetilde{g}_{x} ) \cdot \big( p + \widetilde{A}(x) \big) + \ii \Phi ( \ii e_{x} ).
\end{equation*}
Using Lemma \ref{a22}, Estimate \eqref{a30} easily follows. In the case of $H$, the proof is similar. The fact that $\chi ( H^\# ) \mathcal{D} ( N ) \subset \mathcal{D} ( N )$ is then a consequence of \cite[Theorem 6.2.10]{AmBoGe96_01}.
\end{proof}

\begin{lemma}\sl \label{a31}
Let $H^\#$ denote either $H$ or $\widetilde{H}$. For all $n \in \mathbb{N}$ and $z \in \mathbb{C}$, $0 < \pm \im z \leq 1$, the operator $\< x \>^{-n} ( H^\# - z )^{-1} \< x \>^n$ defined on $\D(\< x \>^n)$ extends by continuity to a bounded operator on $\mathcal{H}$  satisfying
\begin{align}\label{e8}
\big\Vert \< x \>^{-n} ( H^\# - z )^{-1} \< x \>^n \big\Vert \leq \frac{\mathrm{C}}{\vert\im z \vert^{n+1}} .
\end{align}
Moreover, $\< x \>^{-n} ( H^\# - z )^{-1} \< x \>^n (H^\#-z)$ defined on $\D( H^\#)$ extends by continuity to a bounded operator on $\mathcal{H}$ satisfying
\begin{align}\label{e9}
\big\Vert \< x \>^{-n} ( H^\# - z )^{-1} \< x \>^n ( H^\# - z) \big\Vert \leq \frac{\mathrm{C}}{\vert\im z \vert^n} .
\end{align}
\end{lemma}

Estimates \eqref{e8}--\eqref{e9} are established in \cite[Lemma A.5]{BoFa11_01} in the case of $H$. Since the proof is the same in the case of $\widetilde{H}$, we do not reproduce it.

\begin{proof}[Proof of Proposition \ref{a26}]
We show the proposition for $\widetilde{H}$, the case of $H$ being similar. Let $ \eta \in \mathrm{C}_0^\infty( ( - \infty , \Sigma ) )$ be such that $\chi \eta = \chi$. Consider $\varphi \in \mathrm{C}_{0}^{\infty} ( \R ; [ 0 , 1 ] )$ equal to $1$ near $0$ and $\varphi_{R} ( \cdot ) = \varphi( \cdot / R )$ for $R > 0$. Let $u \in \D ( \d \Gamma ( \< y \>^{\beta} ) )$. We want to prove that for all $v \in \D ( \d \Gamma ( \< y \>^{\beta} ) )$,
\begin{equation*}
\big\vert \big\< \d \Gamma ( \< y \>^{\beta} ) v , \chi ( \widetilde{H} ) u \big\> \big\vert \leq \mathrm{C}_u \Vert v \Vert.
\end{equation*}
We  write
\begin{align}
\big\vert \big\< \d \Gamma ( \< y \>^{\beta} ) v , \chi ( \widetilde{H} ) u \big\> \big\vert ={}& \lim_{R \to \infty} \big\vert \big\< \d \Gamma \big( \< y \>^{\beta} \varphi_{R} ( y^{2} ) \big) v , \chi ( \widetilde{H} ) \eta ( \widetilde{H} ) u \big\> \big\vert \nonumber \\
\leq{}& \limsup_{R \to \infty} \big\vert \big\< v , \chi ( \widetilde{H} ) \eta ( \widetilde{H} ) \d \Gamma \big( \< y \>^{\beta} \varphi_{R} ( y^{2} ) \big) u \big\> \big\vert  \nonumber \\
&+ \limsup_{R \to \infty} \big\vert \big\< v , \big[ \d \Gamma \big( \< y \>^{\beta} \varphi_{R} ( y^{2} ) \big) , \chi ( \widetilde{H} ) \big] \eta ( \widetilde{H} ) u \big\> \big\vert   \nonumber \\
&+ \limsup_{R \to \infty} \big\vert \big\< v , \chi ( \widetilde{H} ) \big[ \d \Gamma \big( \< y \>^{\beta} \varphi_{R} ( y^{2} ) \big) , \eta ( \widetilde{H} ) \big] u \big\> \big\vert , \label{a32}
\end{align}
where the commutators should be understood in the sense of quadratic forms on $\D ( N )$. By Lemma \ref{a29}, the previous expressions are justified since $\chi ( \widetilde{H} )$ and $\eta ( \widetilde{H} )$ preserve $\D ( N )$. The first term is easily estimated as
\begin{equation} \label{a33}
\big\vert \big\< v , \chi ( \widetilde{H} ) \eta ( \widetilde{H} ) \d \Gamma \big( \< y \>^{\beta} \varphi_{R} ( y^{2} ) \big) u \big\> \big\vert \leq \mathrm{C} \Vert v \Vert \big\Vert \d \Gamma ( \< y \>^{\beta} ) u \big\Vert .
\end{equation}

Let $\widetilde{\chi} \in \mathrm{C}_{0}^{\infty} ( \C )$ denote an almost analytic extension of $\chi$. To estimate the second term of \eqref{a32}, we write
\begin{align}
\big\Vert \big[ \d \Gamma \big( \< y & \>^{\beta} \varphi_{R} ( y^{2} ) \big) , \chi ( \widetilde{H} ) \big] \eta ( \widetilde{H} ) u \big\Vert     \nonumber \\
\leq{}& \frac{1}{\pi} \int \Big\vert \frac{\partial \widetilde{\chi} (z)}{\partial \bar z} \Big\vert \big\Vert \big[ \d \Gamma \big( \< y \>^{\beta} \varphi_{R} ( y^{2} ) \big) , ( \widetilde{H} - z )^{-1} \big] \eta ( \widetilde{H} ) u \big\Vert \ddre z \ddim z \nonumber \\
\leq{}& \frac{1}{\pi} \int \Big\vert \frac{\partial \widetilde{\chi} (z)}{\partial \bar z} \Big\vert \big\Vert ( \widetilde{H} - z )^{-1} B_{R} ( \widetilde{H} - z )^{-1} \eta ( \widetilde{H} ) u \big\Vert \ddre z \ddim z     \nonumber \\
\leq{}& \frac{1}{\pi} \int \Big\vert \frac{\partial \widetilde{\chi} (z)}{\partial \bar z} \Big\vert \big\Vert ( \widetilde{H} - z )^{-1} \< \widetilde{H} \>^{\frac{1}{2}} \big\Vert \big\Vert \< \widetilde{H} \>^{-\frac{1}{2}} B_{R} \big( N + \< x \>^{\frac{4}{\mu} + 2 \beta} \big)^{-1}  \big\Vert \nonumber \\
&\times \big\Vert \big( N + \< x \>^{\frac{4}{\mu} + 2 \beta} \big) ( \widetilde{H} - z )^{-1} \eta ( \widetilde{H} ) (N+1)^{-1} \big\Vert \Vert (N+1) u \Vert \ddre z \ddim z , \label{a34}
\end{align}
where $B_{R}$ is the quadratic form on $\D ( \widetilde{H} ) \cap \D( N )$ defined by
\begin{equation*}
B_{R} : = \big[ \widetilde{H} , \d \Gamma \big( \< y \>^{\beta} \varphi_{R} ( y^{2} ) \big) \big] .
\end{equation*}
Using Lemma \ref{a29}, one verifies that
\begin{equation*}
\big\Vert  N ( \widetilde{H} - z )^{-1} \eta ( \widetilde{H} ) ( N + 1 )^{-1} \big\Vert \lesssim \vert \im z \vert^{-2} ,
\end{equation*}
and by Theorem \ref{a23},
\begin{equation*}
\big\Vert \< x \>^{\frac{4}{\mu} + 2 \beta} ( \widetilde{H} - z )^{-1} \eta ( \widetilde{H} ) \big\Vert \lesssim \vert \im z \vert^{-1} \big\Vert \< x \>^{\frac{4}{\mu} + 2 \beta} \eta ( \widetilde{H} ) \big\Vert \lesssim \vert \im z \vert^{-1}.
\end{equation*}
We claim that
\begin{equation} \label{a35}
\Big\Vert \< \widetilde{H} \>^{-\frac{1}{2}} B_{R} \big( N+\< x \>^{\frac{4}{\mu} + 2 \beta} \big)^{-1} \Big\Vert \lesssim 1 .
\end{equation}
Then \eqref{a34}--\eqref{a35} together with the properties of $\widetilde{\chi}$ imply that
\begin{equation}  \label{a36}
\big\Vert \big[ \d \Gamma \big( \< y \>^{\beta} \varphi_{R} ( y^{2} ) \big) , \chi ( \widetilde{H} ) \big] \eta ( \widetilde{H} ) u \big\Vert \lesssim \big\Vert \big( \d \Gamma ( \< y \>^{\beta} ) + 1 \big) u \big\Vert .
\end{equation}

Let us now prove \eqref{a35}. In the sense of quadratic forms on $\D( \widetilde{H} ) \cap \D( N )$, we have
\begin{align}
B_{R} ={}& \d \Gamma \big( \big[ \vert k \vert , \< y \>^{\beta} \varphi_{R} ( y^{2} ) \big] \big) + \ii \big( p + \widetilde{A} (x) \big) \cdot \Phi \big( \ii \< y \>^{\beta} \varphi_{R} ( y^{2} ) \widetilde{g}_{x} \big) \nonumber \\
&+ \ii \Phi \big( \ii \< y \>^{\beta} \varphi_{R} ( y^{2} ) \widetilde{g}_{x}) \big) \cdot \big( p + \widetilde{A} (x) \big) + \ii \Phi \big( \ii \< y \>^{\beta} \varphi_{R} ( y^{2} ) e_{x} \big) \nonumber \\
={}& \d \Gamma \big( \big[ \vert k \vert , \< y \>^{\beta} \varphi_{R} ( y^{2} ) \big] \big) - \Phi \big( \ii \< y \>^{\beta} \varphi_{R} ( y^{2} ) \nabla_x \widetilde{g}_{x} \big) + \im \big\< \widetilde{g}_{x} , \ii \< y \>^{\beta} \varphi_{R} ( y^{2} ) \widetilde{g}_{x} \big\>  \nonumber \\
&+ 2 \ii \big( p + \widetilde{A} (x) \big) \cdot \Phi \big( \ii \< y \>^{\beta} \varphi_{R} ( y^{2} ) \widetilde{g}_{x} \big) + \ii \Phi \big( \ii \< y \>^{\beta} \varphi_{R} ( y^{2} ) e_{x} \big) .
\end{align}
Using \eqref{a57} and applying Lemma \ref{a3} (with $t=1$), we obtain that, for all $x \in \mathbb{R}^3$,
\begin{equation*}
\big\Vert \< y \>^{\beta} \varphi_{R} ( y^{2} ) \widetilde{g}_{x} ( k , \lambda ) \big\Vert_{\mathrm{L}^2( \mathbb{R}^3 ; \C^{2} )} \leq \big\Vert \< y \>^{\beta} \widetilde{g}_{x} ( k , \lambda ) \big\Vert_{\mathrm{L}^2( \mathbb{R}^3 ; \C^{2} )} \lesssim \< x \>^{\frac{1}{\mu} + \beta} ,
\end{equation*}
and likewise with $\nabla_x \widetilde{g}_{x} ( k , \lambda )$ or $e_{x} ( k , \lambda )$ in place of $\widetilde{g}_{x} ( k , \lambda )$. Therefore, by Lemma \ref{a7},
\begin{gather}
\big\Vert \big\< \widetilde{g}_{x} , \ii \< y \>^{\beta} \varphi_{R} ( y^{2} ) \widetilde{g}_{x} \big\> \< x \>^{- \frac{2}{\mu} - \beta} \big\Vert \lesssim 1 ,    \label{a37}   \\
\big\Vert \Phi \big( \ii \< y \>^{\beta} \varphi_{R} ( y^{2} ) \nabla_x \widetilde{g}_{x} \big) \< x \>^{- \frac{1}{\mu} - \beta} ( N + 1 )^{-\frac{1}{2}} \big\Vert \lesssim 1 ,     \label{a38}   \\
\big\Vert \Phi \big( \ii \< y \>^{\beta} \varphi_{R} ( y^{2} ) e_{x} \big) \< x \>^{- 1 - \beta} ( N + 1 )^{-\frac{1}{2}} \big\Vert \lesssim 1 ,   \label{a39}
\end{gather}
and, since $\Vert \< \widetilde{H} \>^{-\frac{1}{2}} ( p + \widetilde{A} (x) ) \Vert \lesssim 1$,
\begin{equation}   \label{a40}
\big\Vert \< \widetilde{H} \>^{-\frac{1}{2}} \big( p + \widetilde{A} (x) \big) \cdot \Phi \big( \ii \< y \>^{\beta} \varphi_{R} ( y^{2} ) \widetilde{g}_{x} \big) \< x \>^{- \frac{1}{\mu} - \beta} ( N+1 )^{-\frac{1}{2}} \big\Vert \lesssim 1.
\end{equation}
Finally, using the representation formula
\begin{equation*}
\< y \>^{\beta} \varphi_{R} ( y^{2} ) = \frac{1}{\pi} \int \frac{\partial ( \widetilde{\psi} \widetilde{\varphi}_{R} ) (z)}{\partial \bar z} ( y^{2} - z )^{-1} \ddre z \ddim z ,
\end{equation*}
where $\widetilde{\psi}$ (resp. $\widetilde{\varphi}$) is an almost analytic extension of $( \cdot + 1)^{\frac{\beta}{2}} \in S^{\frac{\beta}{2}} ( \R )$ (resp. $\varphi \in \mathrm{C}_{0}^{\infty} ( \R )$), one can verify that
\begin{equation*}
\big\Vert \big[ \vert k \vert , \< y \>^{\beta} \varphi_{R} ( y^{2} ) \big] \big\Vert \lesssim 1 ,
\end{equation*}
and hence, by Lemma \ref{a21},
\begin{equation} \label{a41}
\big\Vert \d \Gamma \big( \big[ \vert k \vert , \< y \>^{\beta} \varphi_{R} ( y^{2} ) \big] \big) ( N + 1 )^{-1} \big\Vert \lesssim 1 .
\end{equation}
Estimates \eqref{a37}--\eqref{a41} together with the fact that $\Vert \< x \>^{\frac{2}{\mu} + \beta} (N+1)^{1/2} u \Vert \lesssim \Vert ( N + \< x \>^{\frac{4}{\mu} + 2 \beta} ) u \Vert$ imply \eqref{a35}.

It remains to estimate the third term in the right hand side of \eqref{a32}. To this end, let $\widetilde{\eta}$ denote an almost analytic extension of $\eta$ and write similarly
\begin{align}
\big\Vert \chi & ( \widetilde{H} ) \big[ \d \Gamma \big( \< y \>^{\beta} \varphi_{R} ( y^{2} ) \big) , \eta ( \widetilde{H} ) \big] u \big\vert    \nonumber  \\
&\leq \frac{1}{\pi} \int \Big\vert \frac{ \partial \widetilde{\eta} (z) }{ \partial \bar z } \Big\vert \big\Vert \chi( \widetilde{H} ) ( \widetilde{H} - z )^{-1} B_{R} ( \widetilde{H} - z )^{-1} u \big\Vert \ddre z \ddim z   \nonumber \\
&\leq \frac{1}{\pi} \int \Big\vert \frac{ \partial \widetilde{\eta} (z) }{ \partial \bar z } \Big\vert \big\Vert \chi ( \widetilde{H} ) \< x \>^{\frac{2}{\mu} + \beta} \big\Vert \big\Vert \< x \>^{- \frac{2}{\mu} - \beta} ( \widetilde{H} - z )^{-1} \< x \>^{\frac{2}{\mu} + \beta} \< \widetilde{H} \>^{\frac{1}{2}} \big\Vert    \nonumber \\
&\qquad \qquad \qquad \times \big\Vert \< \widetilde{H} \>^{-\frac{1}{2}} \< x \>^{- \frac{2}{\mu} - \beta} B_{R} ( N + 1 )^{-1} \big\Vert \big\Vert ( N + 1 ) ( \widetilde{H} - z )^{-1} (N+1)^{-1} \big\Vert   \nonumber \\
&\qquad \qquad \qquad \times  \Vert ( N + 1 ) u \Vert \ddre z \ddim z.
\end{align}
Theorem \ref{a23} gives $\Vert \chi ( \widetilde{H} ) \< x \>^{\frac{2}{\mu} + \beta} \Vert \lesssim 1$, Lemma \ref{a31} yields $\Vert \< x \>^{- \frac{2}{\mu} - \beta} ( \widetilde{H} - z )^{-1} \< x \>^{\frac{2}{\mu} + \beta} \< \widetilde{H} \>^{\frac{1}{2}} \Vert \lesssim \vert \im z \vert^{- \frac{2}{\mu} - \beta - 1}$, and Lemma \ref{a29} implies $\Vert ( N + 1 ) ( \widetilde{H} - z )^{-1} (N+1)^{-1} \Vert \lesssim \vert \im z \vert^{-2}$. Moreover we claim that
\begin{equation}  \label{a42}
\big\Vert \< \widetilde{H} \>^{-\frac{1}{2}} \< x \>^{- \frac{2}{\mu} - \beta} B_{R} ( N + 1 )^{-1} \big\Vert \lesssim 1 .
\end{equation}
To prove \eqref{a42}, it suffices to proceed in the same way as for \eqref{a35}. The only difference is \eqref{a40}, which is replaced by
\begin{align*}
\Big\Vert \< \widetilde{H} \>^{-\frac{1}{2}} & \< x \>^{- \frac{1}{\mu} - \beta} \big( p + \widetilde{A} (x) \big) \cdot \Phi \big( \ii \< y \>^{\beta} \varphi_{R} ( y^{2} ) \widetilde{g}_{x} \big) ( N+1 )^{-\frac{1}{2}} \Big\Vert    \\
&\leq  \Big\Vert \< \widetilde{H} \>^{-\frac{1}{2}} \< x \>^{- \frac{1}{\mu} - \beta}  \big( p + \widetilde{A} (x) \big) \< x \>^{\frac{1}{\mu} + \beta} \Big\Vert \Big\Vert \< x \>^{- \frac{1}{\mu} - \beta} \Phi \big( \ii \< y \>^{\beta} \varphi_{R} ( y^{2} ) \widetilde{g}_{x} \big) ( N+1 )^{-\frac{1}{2}} \Big\Vert \nonumber \\
&\lesssim \big\Vert \< \widetilde{H} \>^{-\frac{1}{2}} \big( p + \widetilde{A} (x) \big) \big\Vert  + \Big\Vert \< \widetilde{H} \>^{-\frac{1}{2}} \< x \>^{-1} \frac{ \ii x }{ \< x \> } \Big\Vert \lesssim 1.
\end{align*}
Therefore
\begin{equation}\label{a43}
\big\Vert \chi ( \widetilde{H} ) \big[ \d \Gamma \big( \< y \>^{\beta} \varphi_{R} ( y^{2} ) \big) , \eta ( \widetilde{H} ) \big] u \big\Vert \lesssim \big\Vert \big( \d \Gamma ( \< y \>^{\beta} ) + 1 \big) u \big\Vert .
\end{equation}
Equation \eqref{a32} together with the estimates \eqref{a33}, \eqref{a36} and \eqref{a43} conclude the proof of the proposition.
\end{proof}

\section{Creation and annihilation operators} \label{s8}

Let $ \fh:= \mathrm{L}^2 ( \R^3_{k} ; \C^2)$ be the Hilbert space of a photon. The variable $k\in\R^3$ is the wave vector or momentum of the particle. Recall that the propagation speed of the light and the Planck constant divided by $2 \pi$ are set equal to 1. The Bosonic Fock space, $\cF$, over $\fh$ is defined by
\begin{equation*}
\cF : = \bigoplus_{n=0}^{\infty} S_n \fh^{\otimes n} ,
\end{equation*}
where $S_n$ is the orthogonal projection onto the subspace of
totally symmetric $n$-particle wave functions contained in the
$n$-fold tensor product $\fh^{\otimes n}$ of $\fh$ and $S_0
\fh^{\otimes 0} := \C $. The vector $\Om:= (1, 0, \ldots )$ is called the \emph{vacuum vector} in
$\cF$. Vectors $\Psi\in \cF$ can be identified with sequences
$(\psi_n)^{\infty}_{n=0}$ of $n$-particle wave functions
$\psi_n(k_1, \lambda_1, \ldots, k_n, \lambda_n)$, where $\lambda_j \in \{1, 2\}$ are the
polarization variables,  which are
totally symmetric in their $n$ arguments, and $\psi_0\in\C$.

The scalar product of two vectors $\Psi$ and $\Phi$ is given by
\begin{equation} \label{F-scalprod}
\langle \Psi , \Phi \rangle := \sum_{n=0}^{\infty} \sum_{\lambda_1, \ldots, \lambda_n} \int \overline{\psi_n (k_1, \lambda_1, \ldots, k_n, \lambda_n)} \varphi_n (k_1, \lambda_1, \ldots, k_n, \lambda_n) \prod^n_{j=1} \d k_j .
\end{equation}

Given a one particle dispersion relation $\omega(k)$, the energy of
a configuration of $n$ \emph{non-interacting} field particles with
wave vectors $k_1, \ldots,k_n$ is given by $\sum^{n}_{j=1}
\omega(k_j)$. We define the \emph{free-field Hamiltonian}, $H_f$,
giving the field dynamics, by
\begin{equation*}
(H_f \Psi)_n(k_1, \lambda_1, \ldots, k_n, \lambda_n) = \Big( \sum_{j=1}^n \omega(k_j)
\Big) \psi_n (k_1, \lambda_1, \ldots, k_n, \lambda_n) ,
\end{equation*}
for $n\ge1$ and $(H_f \Psi)_n =0$ for $n=0$. Here
$\Psi=(\psi_n)_{n=0}^{\infty}$ (to be sure that the right hand side makes
sense, we can assume that $\psi_n=0$, except for finitely many $n$,
for which $\psi_n(k_1, \lambda_1, \ldots, k_n, \lambda_n)$ decrease rapidly at infinity).
Clearly, if $\omega ( k ) = \vert k \vert$, the operator  $H_f$ has the single eigenvalue  $0$ with
the eigenvector $\Omega$ and the rest of the spectrum absolutely
continuous.

With each function $\varphi \in \mathrm{L}^2 ( \R^3 ; \mathbb{C}^2 )$ one
associates an \emph{annihilation operator} $a(\varphi)$ defined as
follows. For $\Psi=(\psi_n)^{\infty}_{n=0}\in \cF$ with the property
that $\psi_n=0$, for all but finitely many $n$, the vector
$a(\varphi) \Psi$ is defined  by
\begin{equation*}
(a(\varphi) \Psi)_n (k_1, \lambda_1, \ldots, k_n, \lambda_n) : = \sqrt{n+1} \sum_\lam \int
\overline{\varphi(k, \lam)} \psi_{n+1}(k, \lam, k_1, \lambda_1, \ldots, k_n, \lambda_n) \, \d k ,
\end{equation*}
for $n\ge 1$ and $(a(\varphi) \Psi)_n=0$ for $n=0$.
These equations define a closable operator $a(\varphi)$ whose
closure is also denoted by $a(\varphi)$. The creation operator $a^*(\varphi)$ is defined to be the adjoint of
$a(\varphi)$ with respect to the scalar product defined in \eqref{F-scalprod}. Since $a(\varphi)$ is anti-linear and
$a^*(\varphi)$ is linear in $\varphi$, we write formally
\begin{equation*}
a(\varphi) = \sum_{\lambda = 1 , 2} \int \overline{\varphi(k, \lam)} a_\lambda(k) \, \d k , \qquad a^*(\varphi) = \sum_{\lambda = 1 , 2} \int \varphi(k, \lam) a_\lambda^*(k) \, \d k ,
\end{equation*}
where $a_\lambda(k)$ and $a_\lambda^*(k)$ are unbounded, operator-valued
distributions. The latter are well-known to obey the \emph{canonical
commutation relations} (CCR):
\begin{equation*}
\big[ a_{\lambda}^{\#}(k) , a_{\lambda'}^{\#}(k') \big] = 0 , \qquad \big[ a_{\lambda}(k) , a_{\lambda'}^*(k') \big] = \delta_{\lambda, \lambda'} \delta (k-k') ,
\end{equation*}
where $a_\lambda^{\#}= a_\lambda$ or $a_\lambda^*$.

Now, using this one can rewrite the quantum Hamiltonian $H_f$ in
terms of the creation and annihilation operators, $a$ and $a^*$, as
\begin{equation*}
H_f = \sum_{\lambda = 1 , 2} \int a_\lambda^*(k) \omega(k) a_\lambda(k) \, \d k ,
\end{equation*}
acting on the Fock space $ \cF$. More generally, for any operator, $t$, on the one-particle space $\mathrm{L}^2 ( \mathbb{R}^3 ; \mathbb{C}^2 )$ we define the operator $\d \Gamma( t )$ on the Fock space $\cF$ by the following formal expression
\begin{equation*}
\d \Gamma( t ) : = \sum_{\lambda = 1 , 2} \int a_\lambda^* ( k ) t a_\lambda ( k ) \, \d k ,
\end{equation*}
where the operator $t$ acts on the $k$-variable
($\d \Gamma( t )$ is the second quantization of $t$). The precise meaning of the latter expression is
\begin{equation*}
\d \Gamma( t )_{\vert S_n \fh^{\otimes n}} = \sum_{j=1}^{n} \underbrace{1 \otimes \cdots \otimes 1 }_{j-1} \otimes t \otimes \underbrace{1 \otimes \cdots \otimes 1 }_{n - j} .
\end{equation*}

\bibliographystyle{amsplain}
\providecommand{\bysame}{\leavevmode\hbox to3em{\hrulefill}\thinspace}
\providecommand{\MR}{\relax\ifhmode\unskip\space\fi MR }
\providecommand{\MRhref}[2]{%
  \href{http://www.ams.org/mathscinet-getitem?mr=#1}{#2}
}
\providecommand{\href}[2]{#2}


\end{document}